\documentclass[sigconf,authordraft,review=false]{acmart}
\usepackage{amsfonts}
\usepackage{graphicx}
\usepackage{tabularx}
\usepackage{makecell}
\usepackage{array}
\usepackage{booktabs} 
\usepackage{caption}
\usepackage{amsthm}

\usepackage{amssymb}

\AtBeginDocument{%
  }

\renewcommand\footnotetextcopyrightpermission[1]{}

\acmConference{Preprint}{Under Review}

\begin{document}

\title{AutoP2C: An LLM-Based Agent Framework for Code Repository Generation from Multimodal Content in Academic Papers}

\sloppy

\author{Zijie Lin}
\affiliation{%
  \institution{University of Science and Technology of China}
  \city{Hefei}
  \country{China}}
\email{lzj741@mail.ustc.edu.cn}

\author{Yiqing Shen}
\affiliation{%
  \institution{Johns Hopkins University}
  \city{Baltimore}
  \country{USA}}
\email{yiqingshen1@gmail.com}

\author{Qilin Cai}
\affiliation{%
  \institution{University of Science and Technology of China}
  \city{Hefei}
  \country{China}}
\email{cql111@mail.ustc.edu.cn}

\author{He Sun}
\affiliation{%
  \institution{University of Science and Technology of China}
  \city{Hefei}
  \country{China}}
\email{hesun@mail.ustc.edu.cn}

\author{Jinrui Zhou}
\affiliation{%
  \institution{University of Science and Technology of China}
  \city{Hefei}
  \country{China}}
\email{zzkevin@@mail.ustc.edu.cn}

\author{Mingjun Xiao}
\affiliation{%
  \institution{University of Science and Technology of China}
  \city{Hefei}
  \country{China}}
\email{xiaomj@ustc.edu.cn}

\begin{abstract}
Machine Learning (ML) research is spread through academic papers featuring rich multimodal content, including text, diagrams, and tabular results.
However, translating these multimodal elements into executable code remains a challenging and time-consuming process that requires substantial ML expertise. 
We introduce ``Paper-to-Code'' (P2C), a novel task that transforms the multimodal content of scientific publications into fully executable code repositories, which extends beyond the existing formulation of code generation that merely converts textual descriptions into isolated code snippets.
To automate the P2C process, we propose AutoP2C, a multi-agent framework based on large language models that processes both textual and visual content from research papers to generate complete code repositories.  
Specifically, AutoP2C contains four stages: (1) repository blueprint extraction from established codebases, (2) multimodal content parsing that integrates information from text, equations, and figures, (3) hierarchical task decomposition for structured code generation, and (4) iterative feedback-driven debugging to ensure functionality and performance. 
Evaluation on a benchmark of eight research papers demonstrates the effectiveness of AutoP2C, which can successfully generate executable code repositories for all eight papers, while OpenAI-o1 or DeepSeek-R1 can only produce runnable code for one paper. 
The code is available at https://github.com/shoushouyu/Automated-Paper-to-Code.
\vspace{-0.1in}
\end{abstract}

\ccsdesc[500]{Computing methodologies~Multi-agent systems}
\ccsdesc[500]{Human-centered computing~Human computer interaction (HCI)
\vspace{-0.1in}}

\keywords{Multimodality, Multi-agent framework, Large Language Model, Code Repository Generation
\vspace{-0.1in}}

\maketitle

\section{Introduction}
Scientific advancements in the Machine Learning (ML) community are typically disseminated via academic papers \cite{hanson2024strain, ayanwale2024analyzing, li2022identifying}.  
Automating the conversion of these papers into code repositories accelerates both the reproducibility and the real-world application of scientific findings \cite{ball2023ai, puangjaktha2024paper, trofimova2024coderefine}. 
As we know, these research papers generally encompass diverse multimodal content, such as textual descriptions, architectural diagrams, and tabular data \cite{song2025multidisciplinary, pramanick2024spiqa, blecher2023nougat}. Multimodal content can enhance human comprehension and can also help generate corresponding code repositories. For example, visual diagrams help design complex model architectures, textual descriptions deepen the understanding of algorithmic implementations, and tabular hyperparameters can be used to construct configuration files.

\begin{figure}[t!]
    \centering
    \resizebox{\linewidth}{!}{\includegraphics{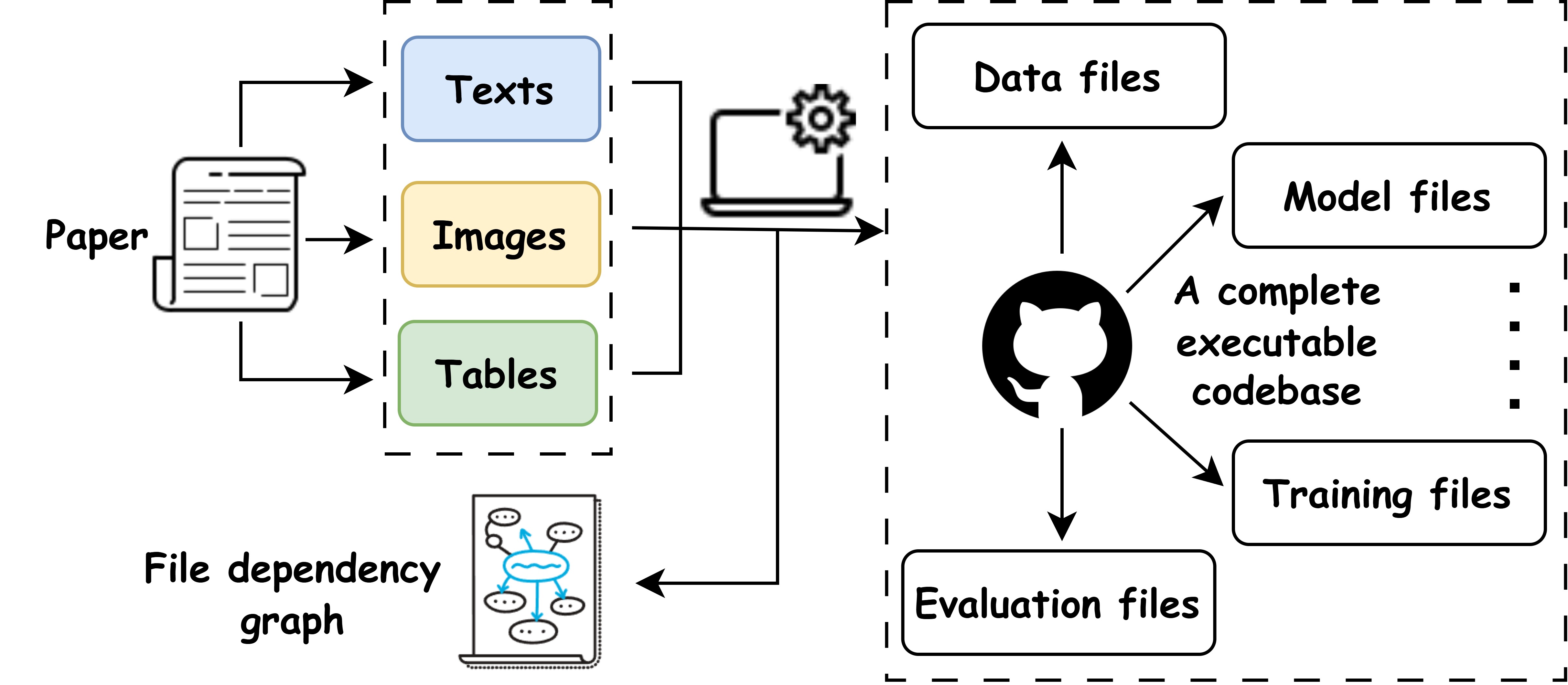}}
    \captionsetup{justification=raggedright} 
    \caption{Illustration of the ``Paper-to-Code'' (P2C) task. 
    It involves processing multimodal content, such as text, images, and tables, from an academic paper to automatically generate a complete, executable code repository together with explanatory diagrams.}
    \vspace{-0.1in}
    \label{fig:task}
\end{figure}

In this paper, we focus on how to automatically generate a code repository for an academic paper with multimodal content and call it the ``Paper-to-Code'' (P2C) task, as depicted in Fig.~\ref{fig:task}.
Specifically, given a research paper containing heterogeneous, multimodal content, P2C aims to generate a complete executable code repository that accurately implements the described methods and reproduces the reported results.
However, unlike traditional code generation, there are three novel challenges in dealing with the P2C task. First, P2C requires processing and integrating of multimodal inputs from academic papers (\textit{e}.\textit{g}., text, images, tables). Second, new model designs must be implemented accurately and align with the paper's descriptions. Third, it needs to generate multi-file, structured code repositories rather than isolated code snippets \cite{sundar2024cpapers, yang2023deep}.

Existing approaches for automated ML implementation face various limitations when handling the P2C task.
First, some related works involve traditional Automated Machine Learning (AutoML) methods \cite{ali2006learning, bergstra2012random, jin2019auto, bisong2019google, das2020amazon}, which are employed to automate the end-to-end process of training machine learning models for structured data. However, they rely on predefined workflows and model pools, rendering them unsuitable for implementing new architectures or unconventional algorithms.  
Second, Large Language Models (LLMs) have been applied to significantly enhance the flexibility of code generation from natural language in recent years \cite{liu2023your, mu2024clarifygpt}. Nonetheless, on the one hand, even state-of-the-art models struggle to generate multi-file code repositories, constrained by context window limitations and inference complexity \cite{anil2022exploring, newman2020eos}.
On the other hand, these LLMs primarily process textual descriptions, neglecting the rich visual and structural multimodal information in research papers.
Third, multi-agent frameworks \cite{yang2024swe, tao2024magis} have emerged as a promising approach for complex code generation by employing specialized agents collaborating in iterative development cycles.
For example, self-organized agents \cite{ishibashi2024self} use a multi-agent system that automatically replicates agents based on task complexity, thereby enabling scalable large-scale code generation. 
However, these multi-agent methods predominantly focus on generic software development scenarios with text-only inputs, lacking the multimodal understanding required to interpret architectural diagrams, parse mathematical equations, and extract implementation details from tables and figures \cite{li2024mmcode}.
As a result, they fall short of providing in-depth research and practical capability for generating fully replicable code repositories from ML research papers.
To fill this research-to-implementation gap, we propose AutoP2C, a multi-agent framework specifically designed to generate complete code repositories from the multimodal content of research papers. 
AutoP2C comprises four stages. 
In the first stage, it extracts universal code structures from established repositories to construct architectural blueprints. 
The next stage involves multimodal content parsing, which integrates information from text, diagrams, and tables into a unified representation.
In the third stage, divide-and-conquer task planning decomposes complex implementations into hierarchical subtasks with clearly defined interfaces. 
Finally, execution feedback-driven debugging localizes errors and aligns the code with the multimodal specifications of the paper through iterative testing.
Unlike previous approaches that treat code generation as a unimodal text-to-text translation problem \cite{trofimova2024coderefine, islam2024mapcoder}, AutoP2C takes advantage of multimodal understanding to capture the full spectrum of information presented in academic papers. 
To evaluate the performance of P2C, we propose two novel metrics to assess the structural completeness of code repositories and combine these with actual runtime performance to quantify how faithfully the generated code reproduces the paper's reported performance.

The major contributions are threefold.
(1) 
We introduce a novel task, \textit{i}.\textit{e}., P2C, and first formalize it as transforming the multimodal content of academic papers into structured, multifile code repositories with explanatory visualizations.
(2) We propose AutoP2C, an innovative multi-agent framework that consists of four carefully designed stages to automate the P2C task. 
Furthermore, AutoP2C produces internal file dependency graphs that facilitate user understanding and review to improve transparency. 
(3) We establish a benchmark spanning multiple ML domains to evaluate P2C. 
Correspondingly, we propose two P2C evaluation metrics for structural completeness and combine them with run-time performance to serve as the metrics for the generated code repositories.

\section{Related Works}

\subsection{Large Language Models for Code Generation}
LLM have demonstrated their capabilities in code generation tasks, evolving from simple code completion to handling increasingly complex programming challenges \cite{nijkamp2022codegen, wang2023codet5+, thakur2024verigen, le2022coderl, dong2024self}.
Self-planning approaches \cite{jiang2024self} have enhanced these LLMs by guiding them to outline solution steps based on user intent before generating code, resulting in more targeted implementations. 
Models like CodeGen \cite{nijkamp2022codegen} have pioneered multistep program synthesis that improves the understanding of user requirements, while DeepSeek-Coder \cite{guo2024deepseek} employs a project-level code corpus with innovative ``fill-in-the-blank'' pre-training objectives to enhance code infill capabilities.
Despite these advances, current LLMs for code generation operate only within a unimodal paradigm \textit{i}.\textit{e.}, processing only textual input \cite{xu2024large, anil2022exploring, zhang2023planning}, which prevents them from effectively interpreting the rich multimodal content found in research papers, such as architectural diagrams, mathematical equations, and tabular results.
Furthermore, these models typically generate isolated code snippets rather than cohesive, multi-file repositories that researchers require for implementation.

\subsection{Multi-Agent for Code Generation}
Single LLMs often struggle with context management, modular design, error detection, and complex reasoning in multiple files in code generation.
Multi-agent frameworks address these shortcomings by distributing specialized tasks across collaborative agents that can maintain focused context awareness while collectively managing repository-level complexity \cite{huang2023agentcoder, wu2023autogen, chen2023autoagents, qian2023communicative}. 
For example, CodeCoR \cite{Pan2025} implements a self-reflective multi-agent where agents collaboratively generate code, develop test cases, and perform iterative repairs to improve both quality and executability. 
Similarly, MetaGPT \cite{Hong2024} adopts a meta-programming paradigm that assigns distinct roles to specialized agents, enabling the coordinated production of coherent and maintainable multi-file code repositories through structured workflows.
Among these frameworks, CodeAgent \cite{zhang2024codeagent} represents the first framework specifically designed for repository-level code generation. 
This LLM-based agent integrates five programming tools through four distinct agent strategies to manage complex code repositories effectively. 
Despite these notable advances, efficiently generating high-quality, multi-file code repositories remains challenging, particularly for implementing ML algorithms described in research papers as they require the understanding for the multimodal context.

\begin{figure*}[t!]
    \centering
    \includegraphics[width=0.9\linewidth]{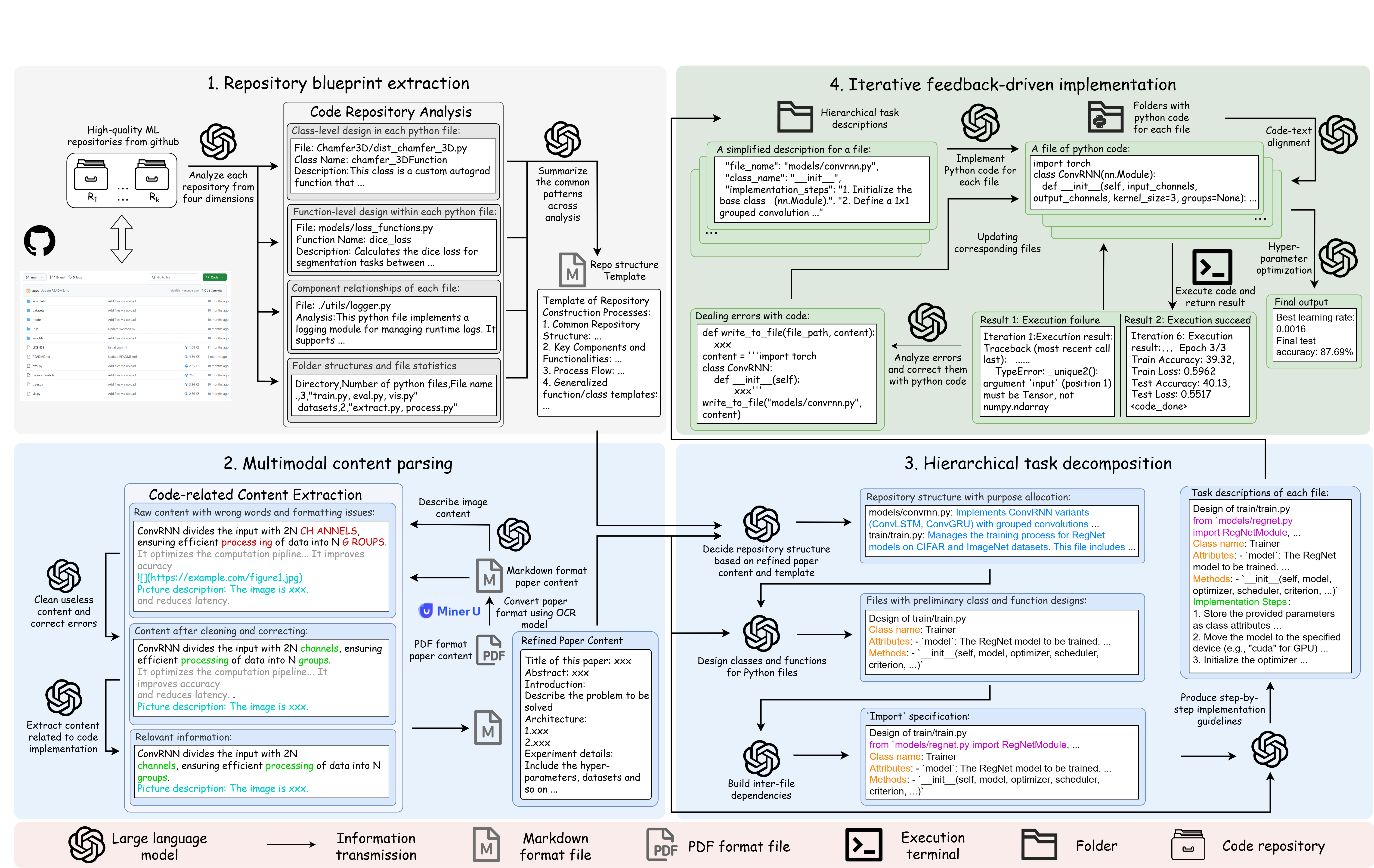}
    \vspace{-0.1in}
    \caption{Overview of the AutoP2C framework. 
    It contains four stage, namely (1) repository blueprint extraction, which analyzes established ML repositories to generate a standardized blueprint for future code organization; (2) multimodal content parsing, which processes PDF papers into a unified, implementation-focused representation; (3) hierarchical task decomposition, which generates structured implementation plans; and (4) iterative feedback-driven implementation, which transforms task descriptions into an executable code repository.
    \vspace{-0.1in}
    }
    \label{fig:framework}
\end{figure*}

\section{Methods}
\subsection{Problem Formulation}
The task of ``Paper-to-Code'' (P2C) is defined as the generation of executable code repositories based on multimodal content from academic papers on ML. 
Formally, we denote the multimodal content in the academic paper $P$ as a triple comprising three modalities, that is, $P \triangleq (\text{Texts}, \text{Images}, \text{Tables})$, 
where $\text{Texts} = \{\text{Paragraph}_1, \ldots, \text{Paragraph}_N\}$ contains all textual content, such as abstract, method, and results; $\text{Images} = \{\text{Image}_1, \ldots, \text{Image}_A\}$ represents visual elements, including architectural diagrams and results visualizations; and $\text{Tables} = \{\text{Table}_1, \ldots, \text{Table}_B\}$ contains structured data, \textit{e}.\textit{g}., experimental results, hyperparameter configurations, and comparative analyzes presented in tabular format.
The target output of P2C is a code repository $R \triangleq \big( \{\text{Module}_1, \dots, \text{Module}_k, \dots, \text{Module}_K\},\ D \big)$, where $\text{Module}_k$ represents a component of the implementation (\textit{e}.\textit{g}., data processor, model architecture, training procedures, evaluation metrics) and it actually is a collection of related source code files, \textit{i}.\textit{e}., $\text{Module}_k \triangleq \{\text{File}_{k,1}, \text{File}_{k,2}, \dots, \text{File}_{k,N_k}\}$.
Here, $N_k \geq 1$ ensures that each $k$-th module contains at least one implementation file. 
Correspondingly, $D$ refers to the generated explanatory diagrams, such as dependency graphs, which further improve the understanding of the repository's structure and facilitate navigation through the generated code repository.
Based on the above definitions, the P2C task can be formalized as
\begin{equation}
    f_{\,\text{P2C}}: P \rightarrow R.
\end{equation}
Importantly, this formulation distinguishes P2C from conventional code generation tasks \cite{lu2021codexglue, austin2021program, gong2024evaluation, labash2024res} in three aspects: 
(1) P2C processes multimodal context rather than purely textual descriptions for code generation; 
(2) P2C produces structured multi-file code repositories instead of isolated code snippets;
(3) P2C generates explanatory multimodal visualizations that improve code comprehensibility.

\begin{figure*}[t!]
    \centering
    \resizebox{0.9\linewidth}{!}{\includegraphics{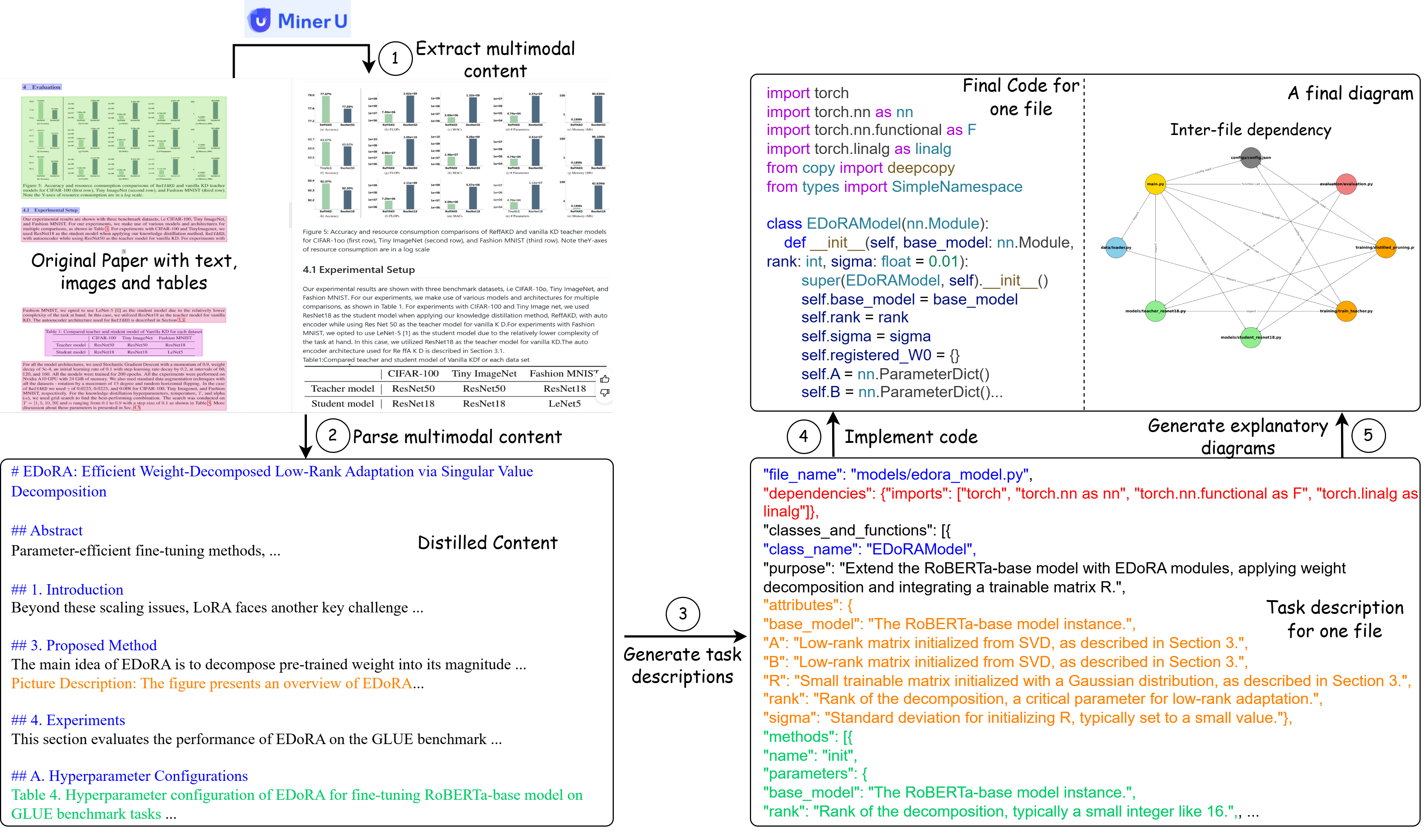}}
    \vspace{-0.1in}
    \caption{Case study illustrates the complete workflow of AutoP2C for the EDoRA: ... \cite{nasiri2025edora} paper. 
    The process begins with (1) extracting multimodal content (text, images, tables) from the original research paper using MinerU \cite{wang2024mineru}, followed by (2) multimodal content parsing to produce refined, structured content with cross-modal relationships preserved. 
    Based on this processed multimodal information, AutoP2C then performs (3) hierarchical task decomposition to generate detailed file descriptions with explicit references to paper sections.
    These descriptions guide (4) the implementation of functional code that accurately captures the paper's algorithmic specifications, culminating in (5) the generation of explanatory diagrams that visualize component relationships in the resulting code repository. }
    \vspace{-0.1in}
    \label{fig:wide_example2}
\end{figure*}

\subsection{Method Overview}
To address the P2C task, we propose AutoP2C, a multi-agent framework based on LLM shown in Fig.~\ref{fig:framework}. 
AutoP2C consists of four stages that imitate human experts implementing code repositories from research papers.
The first stage is the repository blueprints extraction, where an LLM analyzes well-established ML code repositories of different domains $\mathcal{R} = \{R_1, R_2, \dots, R_m\}$ to extract common architectural patterns as the ``base knowledge''.
Specifically, this stage produces a standardized repository organization template $T$ across four dimensions, namely repository architecture, file interdependencies, function designs, and class structures.
This repository-level understanding provides scaffolding for subsequent code generation and is performed only once as a pre-processing step.

The second stage, multimodal content parsing, processes academic papers from PDF format to extract necessary information across modalities for code implementation. 
This stage employs the OCR model \cite{wei2024general, chen2025ocean} for textual content extraction, the Vision Language Model (VLM) \cite{wu2024visionllm, chu2024mobilevlm} for the understanding of architectural diagrams, the LLM for the analysis of mathematical equations for algorithmic formulations and the extraction of structured data for configurations and results.
Afterward, an LLM synthesizes all these diverse elements from different modalities into a unified representation $P_{\text{distilled}}$ in text that preserves the major cross-modal relationships necessary for code implementation.

With the content of the distilled paper $P_{\text{distilled}}$ purely in text together with the repository blueprint $T$ at hand, the hierarchical task decomposition stage uses Large Reasoning Models (LRMs) \cite{guo2025deepseek} to generate a structured implementation plan with task descriptions $\mathcal{T} = \{\tau_1, \tau_2, \dots, \tau_n\}$ for each component, defining file functionalities, dependencies, and interfaces while ensuring alignment with the paper's algorithmic specifications.
Concretely, the decomposition of AutoP2C follows a top-down design, beginning with high-level architectural decisions and progressively refining into specific implementation details.

The final stage, namely iterative feedback-driven implementation, transforms task descriptions $\mathcal{T}$ into a fully executable code repository $R_{\text{exec}}$ through an implementation-verification loop. 
Specifically, a coding agent generates initial implementations for each module based on their respective task description. 
Then, a verification agent executes the code, identifies errors or inconsistencies, and provides targeted feedback for refinement.
This feedback drives subsequent refinement iterations until the repository successfully reproduces the results described in the paper. 
Additionally, at this stage, AutoP2C generates multimodal explanatory visualizations that enhance code comprehensibility by illustrating data flows throughout the generated code repository.

\subsection{Repository Blueprint Extraction}
To establish a repository organization framework that accommodates diverse implementation patterns, AutoP2C first derives standardized blueprints from existing high-quality ML code repositories before processing any specific paper.
Given a set of representative code repositories $\mathcal{R} = \{R_1, R_2, \dots, R_m\}$ that span various domains of ML, we employ LLMs to extract structural patterns in four hierarchical dimensions. 
The repositories are specifically selected based on three criteria: (1) popularity metrics in the ML community (\textit{e}.\textit{g}., $>1000$ GitHub stars), (2) code quality indicators (\textit{e}.\textit{g}., with comprehensive documentation), and (3) domain diversity to ensure coverage across computer vision, natural language processing, graph and others.

First, on the repository level, the LLM examines the overall architecture by quantifying folder organization patterns and file distribution statistics, denoted as $A_{\text{arch}}(R_i) = \{(\text{folder}_i^n, |F_i^n|, F_i^n = \{f^{n,1}_i, f^{n,2}_i, \dots, f^{n,j}_i, \dots\})| R_i \in \mathcal{R}, n \ge 1\}$,
where $\text{folder}_i^n$ represents a top-level folder (\textit{e}.\textit{g}., ``models'', ``data'', ``utils'') in $R_i$, $f_i^{n,j}$ denotes the individual files contained within that folder, and $|F_i^n|$ quantifies the number of files.

This analysis is performed through automated repository structure parsing combined with LLM-guided categorization that classifies folders according to their functional purpose. 
For instance, the LLM recognizes that folders containing predominantly class definitions represent model architectures, while those with data-loading operations represent preprocessing components.

Second, on the folder level, the LLM determines the relationships between folders by extracting the interdependencies of files in these folders and the workflow sequence for each file in repository $R_i$, denoted as $A_{\text{relationship}}(R_i) = \{(f_i^j, D_i^j, O_i^j)| R_i \in \mathcal{R}, j \ge 1\}$, where for each file $f_i^j$, $D_i^j$ is its dependency graph encoded as a structured list of import statements and function calls to other modules, and $O_i^j$ captures its operational logic as a summarized text representation of the file's primary functionality. Here, for simplicity, we ignore the folder information in $f_i^j$.

Third, on the function level, the LLM characterizes design patterns of functions in each file, denoted as $A_{\text{func\_design}}(R_i) = \{(\text{func}_{i,j}^a, \text{sig}_{i,j}^a, \text{flow}_{i,j}^a)| R_i \in \mathcal{R}, a \ge 1\}$, where $\text{func}_{i,j}^a$ is the \textit{a}-th function of the \textit{j}-th file in repository $R_i$, represented by its signature $\text{sig}_{i,j}^a$ (including parameter types, default values, and return specifications) and execution flow $\text{flow}_{i,j}^a$ (a structured representation of control flow including conditional branches and loop structures). 

Fourth, on the class level, the LLM identifies design patterns of classes that depict domain-specific concepts, denoted as $A_{\text{class\_design}}(R_i) = \{(\text{class}_{i,j}^b, \text{attr}_{i,j}^b, \text{method}_{i,j}^b)|R_i \in \mathcal{R}, b \ge 1\}$, where $\text{class}_{i,j}^b$ is the \textit{b}-th class of the \textit{j}-th file in repository $R_i$, characterized by its attributes (\textit{i}.\textit{e}., $\text{attr}_{i,j}^b$), which define the class's state, such as configuration parameters, model weights, or dataset properties, and methods (\textit{i}.\textit{e}., $\text{method}_{i,j}^b$) that define the class's behavior, including initialization, forward passes, and utility functions. 

After analyzing individual repositories, the repository blueprint extraction stage synthesizes a unified template by identifying common patterns across repositories,
\begin{equation}
\begin{aligned}
T =& \text{LLM}_\text{synthesize}(\{A_{\text{arch}}(R_i), A_{\text{relationship}}(R_i), \\
&A_{\text{func\_design}}(R_i), A_{\text{class\_design}}(R_i)\}_{i=1}^m).
\end{aligned}
\end{equation} 
The synthesize operation employs LLM to identify recurring structures across multiple repositories. 
This process involves frequency analysis to determine common directory structures (\textit{e}.\textit{g}., data/models/utils pattern), abstraction of implementation details to create generalized function templates, and identification of standard interface conventions between components.
The resulting template $T$ is represented as a hierarchical markdown structure with four main sections that correspond to the above analyzed dimensions.

\subsection{Multimodal Content Parsing}

The multimodal parsing process begins with transforming the PDF format of paper $P$ into a markdown format by using an OCR model MinerU \cite{wang2024mineru}.
This initial conversion yields a raw representation $P_{\text{raw}}= \text{OCR}_\text{MinerU}(P) = (T_{\text{raw}}, I_{\text{raw}}, M_{\text{raw}}, D_{\text{raw}})$, where $T_{\text{raw}}$ represents textual content that includes methodology descriptions and algorithmic procedures; $I_{\text{raw}}$ contains visual elements such as architectural diagrams, flow charts, and result visualizations; $M_{\text{raw}}$ contains mathematical equations; and $D_{\text{raw}}$ is tabular data such as hyperparameter configurations and experimental results. 
However, this raw conversion typically introduces artifacts such as fragmented paragraphs, misplaced image references, and misformed equations that must be addressed before future content extraction \cite{liu2024ocrbench}. 
To restore structural integrity, AutoP2C employs an LLM to identify and correct these conversion artifacts, \textit{i}.\textit{e}. $T_{\text{structured}} = \text{LLM}_\text{restore}(T_{\text{raw}})$.

Next, AutoP2C tries to integrate multimodal content.
For visual content processing, AutoP2C employs VLMs \cite{wu2024visionllm, chu2024mobilevlm} to extract semantic understanding from visual elements to generate structured descriptions that capture component relationships, depicted as $I_{\text{parsed}} = \{\text{Parse}_{\text{visual}}(i) | i \in I_{\text{raw}}\}$.
Rather than relying on VLMs to directly describe the images, we design a guiding prompt consisting of three key components: (1) Emphasizing only code-relevant information by disregarding superfluous visual details (e.g., color), so as to avoid introducing noise and distracting attention. (2) Providing a comprehensive description, particularly including all numerical elements in the image to prevent omission. (3) Leveraging the image’s caption to establish strong correlations between the image and relevant text from the paper, thereby producing an accurate, text-aligned answer to minimize ambiguity.
Similarly, mathematical equations are translated from typeset representations into computational forms via LLM \textit{i}.\textit{e}., $M_{\text{parsed}} = \{\text{Parse}_{\text{math}}(m) | m \in M_{\text{raw}}\}$.
Tabular data undergoes similar extraction to identify hyperparameters, experimental configurations, and evaluation metrics by
$D_{\text{parsed}} = \{\text{Parse}_{\text{tabular}}(d) | d \in D_{\text{raw}}\}$.

Finally, after processing individual modalities, AutoP2C integrates with LLM to remove redundancy, namely
$P_{\text{integrated}} = \text{LLM}_\text{integrate}(T_{\text{structured}}, I_{\text{parsed}}, M_{\text{parsed}}, D_{\text{parsed}})$.
The final stage of multimodal parsing filters the integrated content to focus on code implementation-relevant information with LLM, namely
$P_{\text{distilled}} = \text{LLM}_{\text{filter}}(P_{\text{integrated}})$.
This filtering process identifies and retains content directly relevant to code implementation, including algorithmic descriptions, architectural specifications, hyperparameter settings, and evaluation methodologies, while deprioritizing content such as literature reviews, theoretical justifications, and references that contribute less directly to implementation.

\subsection{Hierarchical Task Decomposition}
With the distilled multimodal content $P_{\text{distilled}}$ and the repository blueprint $T$, AutoP2C progresses to decomposing the implementation into hierarchically structured tasks with LLM. 
The decomposition process follows a top-down design that progressively refines implementation requirements across multiple levels of abstraction. 
First, AutoP2C generates a repository-level architecture that defines the overall structure with LLM \textit{i}.\textit{e}.,
$A_{\text{repo}} = \text{LLM}_\text{architecture}(P_{\text{distilled}}, T) = \{(F_i, \phi_i, S_i)\}_{i=1}^n$,
where $F_i$ indicates a file in the repository and is associated with its description of the functionality $\phi_i$ and the corresponding source sections $S_i$ of the academic paper $P_{\text{distilled}}$. 
It uses the repository blueprint $T$ to ensure that the generated architecture adheres to established best practices while accommodating the unique contributions of the article.

Second, AutoP2C generates detailed component specifications for each file in $A_{\text{repo}}$, namely
$C_i = \text{LLM}_\text{components}(F_i, P_{\text{distilled}}) = \{(\kappa_j, \alpha_j, \mu_j)\}_{j=1}^{m_i}$,
where $\kappa_j$ represents a class or function, $\alpha_j$ defines its attributes or parameters, $\mu_j$ specifies its methods or behaviors, and $m_i$ is the number of classes or functions in $F_i$. 
To prevent semantic drift \cite{ji2023survey} \textit{i}.\textit{e}, the gradual deviation from the original algorithmic intent during implementation, each component is explicitly linked to specific sections of the paper that describe its algorithmic or architectural details.
Meanwhile, AutoP2C constructs explicit modeling of inter-file dependencies, namely
$D = \text{LLM}_\text{MapDependencies}({C_i}_{i=1}^n) = \{(F_i, \{(F_j, \epsilon_{i,j})\})\}$
where $\epsilon_{i,j}$ represents the specific components from file $F_j$ upon which file $F_i$ depends. 
This dependency mapping ensures structural coherence across the code repository and facilitates proper interface design between components. 
Importantly, AutoP2C prioritizes internal dependencies over external library dependencies.

Third, to mitigate the complexity of reasoning for LLMs when designing intricate implementation details, AutoP2C employs an iterative refinement approach to generate task descriptions for each file, denoted as
$\tau_i = \text{LLM}_\text{Gen}(F_i, C_i, D, P_{\text{distilled}})$.
This process focuses on one file at a time, refining its implementation details while maintaining awareness of the global repository structure. 
For each file, AutoP2C produces step-by-step implementation guidelines that explicitly refer to relevant sections of the paper containing algorithmic specifications, mathematical formulations, or architectural diagrams.
Finally, the set of task descriptions $\mathcal{T} = \{\tau_1, \tau_2, ..., \tau_n\}$ serves as a blueprint for implementation.

Additionally, to enhance interpretability and facilitate human review, AutoP2C also generates a visual representation of inter-file dependencies. 
This visualization is created through a graph rendering process that transforms the formal dependency structure $D$ into an interactive diagram. 
The nodes in the graph represent files, while the directed edges represent dependencies, with edge labels indicating the specific nature of each dependency (\textit{e}.\textit{g}., inheritance, import). 
The visualization uses color coding to differentiate between component types (\textit{e}.\textit{g}., models, training).

\subsection{Iterative Feedback-Driven Implementation}
The final stage of AutoP2C transforms hierarchical task descriptions $\mathcal{T} = \{\tau_1, \tau_2, ..., \tau_n\}$ into a fully executable code repository through an iterative implementation-verification process. 
Specifically, AutoP2C employs a progressive file-by-file implementation approach
\begin{equation}
    Code_i = \text{LLM}_\text{ImplementCode}(\tau_i, \{Code_j | j < i\}, P_{\text{distilled}}),
\end{equation}
where each code file $Code_i$ is generated based on its corresponding task description $\tau_i$, previously implemented code files $\{Code_j | j < i\}$, and the distilled paper representation $P_{\text{distilled}}$. 
This sequential approach ensures that the code implementation remains consistent across the repository, with each new file building on the existing implementation context.
The implementation order follows the dependency graph established during the hierarchical task decomposition stage, beginning with foundational components and progressively implementing dependent modules.

Next, AutoP2C performs a validation process to ensure the fidelity of implemented code files to the core algorithms of the paper \textit{i}.\textit{e}., $V = \text{LLM}_\text{validate}(\{Code_i\}_{i=1}^ n, P_\text{distilled})$.
This validation focuses on three aspects: (1) model architecture design, verifying that neural network structures match the paper's specifications; 
(2) loss function formulation, ensuring mathematical correctness in the implementation of objective functions; and (3) optimization strategy, confirming that update rules align with the described training procedures.
The validation process employs cross-modal verification, comparing code implementations with relevant sections of the paper through textual descriptions, architectural diagrams, and mathematical formulations to ensure complete alignment.

Despite the validation step, producing a fully executable pipeline in a single pass remains challenging due to complex inter-file interactions and interface consistency requirements. 
To address this, AutoP2C implements an iterative debugging mechanism with LLM, namely $\{Code_i'\}_{i=1}^n = \text{LLM}_\text{Debug}(\{Code_i\}_{i=1}^n, E)$, where $\{Code_i'\}_{i=1}^n$ represents the refined code and $E$ denotes execution errors encountered during testing. 
The debugging process operates through a two-phase approach.
First, the error localization phase, \textit{i}.\textit{e}., $L = \text{LLM}_\text{LocalizeError}(E, \{Code_i\}_{i=1}^n)$, identifies the specific files and components associated with each LLM execution error, minimizing the scope of necessary modifications and reducing the risk of unintended side effects.
Second, the error correction phase, \textit{i}.\textit{e}., $\{Code_i'\} = \text{LLM}_\text{CorrectError}(L, \{Code_i\}, P_{\text{distilled}})$, generates targeted modifications to resolve identified issues while maintaining adherence to the paper's specifications, which may involve adjusting function interfaces, aligning data structures, or refining implementation details to ensure compatibility across components.

After successful debugging, AutoP2C integrates hyperparameter optimization (HPO) through an automated process following previous work \cite{luo2024autom3l}.
Specifically, we incorporate \texttt{ray.tune} \cite{raiaan2024systematic} into the main execution file, enabling exploration of the hyperparameter space defined in the paper. 
The HPO implementation extracts relevant hyperparameters from the experimental sections of the paper and configures appropriate search spaces based on reported values and ranges.

\renewcommand{\arraystretch}{1.7} 
\setlength{\tabcolsep}{2pt} 
\newcolumntype{P}[1]{>{\centering\arraybackslash}p{#1}}

\begin{table*}[t]
    \centering
    \caption{Comparative evaluation of AutoP2C against baseline models (OpenAI-o1 and DeepSeek-R1) across eight research papers in terms of four metrics, \textit{i}.\textit{e}., absolute performance (accuracy on target datasets), relative performance (ratio to original implementation), class completeness ($\text{COMP}_{\text{class}}$) and function completeness ($\text{COMP}_{\text{func}}$). 
    Red crosses (\textcolor{red}{$\times$}) indicate failure to generate executable code. 
    The rightmost columns show computational resource consumption through input and output token counts for the AutoP2C framework. 
    Results demonstrate AutoP2C's superior ability to translate multimodal paper content into functional code repositories, with improvements in both execution success and structural completeness. 
    }
    \vspace{-0.1in}
    \label{tab:merged}
    \small
    \resizebox{\textwidth}{!}{%
    \begin{tabular}{P{2.6cm}P{1.5cm}P{1.4cm}%
                    P{0.65cm}P{0.65cm}P{0.65cm}%
                    P{0.65cm}P{0.65cm}P{0.65cm}%
                    P{0.65cm}P{0.65cm}P{0.65cm}%
                    P{0.65cm}P{0.65cm}P{0.65cm}%
                    P{1cm}P{1cm}}
        \toprule
        Paper's Title & ArXiv ID & Task Type & \multicolumn{3}{c}{Performance} & \multicolumn{3}{c}{Relative Performance} & \multicolumn{3}{c}{$\text{COMP}_{\text{class}}$} & \multicolumn{3}{c}{$\text{COMP}_{\text{func}}$} & {$\text{Tokens}_{\text{in}}$} & {$\text{Tokens}_{\text{out}}$} \\
        \cmidrule(lr){4-6} \cmidrule(lr){7-9} \cmidrule(lr){10-12} \cmidrule(lr){13-15} \\[-6ex]
         &  &  & o1 & R1 & ours & o1 & R1 & ours & o1 & R1 & ours & o1 & R1 & ours &  &  \\
        \midrule
        Distilled... \cite{fontana2024distilled}        & None       & multiclass & \textcolor{red}{$\times$} & \textcolor{red}{$\times$} & \textbf{82.8\%} & \textcolor{red}{$\times$} & \textcolor{red}{$\times$} & \textbf{89.8\%} & 90.8\%  & 90.8\% & \textbf{90.8\%}  & 31.3\% & 37.8\% & \textbf{60.2\%}  & 852K   & 120K \\
        EDoRA: ... \cite{nasiri2025edora}          & 2501.12067 & binary     & \textcolor{red}{$\times$} & \textcolor{red}{$\times$} & \textbf{82.5\%} & \textcolor{red}{$\times$} & \textcolor{red}{$\times$} & \textbf{89.9\%}  & 27.1\% & 0\%    & \textbf{54.3\%}  & 42.5\% & 17.0\% & \textbf{88.9\%}  & 458K   & 93K  \\
        No More Adam:... \cite{xu2024no}      & 2412.11768 & multiclass & 81.8\%     & 75.3\%     & \textbf{91.4\%} & 87.5\%     & 80.5\%     & \textbf{97.8\%} & 43.1\% & 24.8\% & \textbf{86.1\%}  & None   & None   & None    & 1177K  & 145K \\
        Entropy Aware... \cite{nazari2024entropy}    & 2403.04636 & multiclass & \textcolor{red}{$\times$} & \textcolor{red}{$\times$} & \textbf{79.9\%} & \textcolor{red}{$\times$} & \textcolor{red}{$\times$} & \textbf{103.0\%} & 15.3\% & 3.0\%  & \textbf{53.8\%}  & 9.1\% & 8.0\%  & \textbf{46.2\%}  & 334K   & 77K  \\
        $L^2$GC:... \cite{liang20242gc}         & 2403.06064 & multiclass & \textcolor{red}{$\times$} & \textcolor{red}{$\times$} & \textbf{79.5\%} & \textcolor{red}{$\times$} & \textcolor{red}{$\times$} & \textbf{96.5\%}  & 11.0\% & 50.9\% & \textbf{75.9\%}  & 18.4\% & 12.5\% & \textbf{31.4\%}  & 477K   & 106K \\
        ReffAKD:... \cite{doshi2024reffakd}         & 2404.09886 & multiclass & \textcolor{red}{$\times$} & \textcolor{red}{$\times$} & \textbf{91.6\%} & \textcolor{red}{$\times$} & \textcolor{red}{$\times$} & \textbf{98.0\%}  & 13.0\% & 11.7\% & \textbf{25.9\%}  & 23.8\% & 23.2\% & \textbf{35.9\%}  & 547K   & 112K \\
        KD-LoRA:... \cite{azimi2024kd}         & 2410.20777 & binary     & \textcolor{red}{$\times$} & \textcolor{red}{$\times$} & \textbf{91.1\%} & \textcolor{red}{$\times$} & \textcolor{red}{$\times$} & \textbf{99.0\%}  & None   & None   & None    & 60.0\% & 20.0\% & \textbf{80.0\%}  & 310K   & 88K  \\
        Convolutional... \cite{linse2023convolutional}    & 2411.18388 & multiclass & \textcolor{red}{$\times$} & \textcolor{red}{$\times$} & \textbf{92.0\%} & \textcolor{red}{$\times$} & \textcolor{red}{$\times$} & \textbf{122.0\%}  & 43.7\% & 36.7\% & \textbf{73.1\%}  & 18.5\% & 4.4\%  & \textbf{18.5\%}  & 501K   & 98K  \\
        \bottomrule
    \end{tabular}%
    \vspace{-0.1in}
    }
\end{table*}

\renewcommand{\arraystretch}{1.25} 
\setlength{\tabcolsep}{4pt}       
\newcolumntype{P}[1]{>{\centering\arraybackslash}p{#1}}

\begin{table}[t]
    \centering
    \large
    \caption{Average Replication Scores (in \%) of different baselines on PaperBench Code-Dev benchmark using o3-mini-high. The results of the baseline in the first three lines are all from the paper \cite{seo2025paper2code}. `$\pm$' represents the standard error of the mean.}
    \vspace{-0.1in}
    \label{tab:code_dev_scores}
    \small
    \resizebox{\columnwidth}{!}{%
    \begin{tabular}{P{3cm} P{4cm}}
    \toprule
    \textbf{Baseline} & \textbf{Replication Score (\%)} \\
    \midrule
    
    BasicAgent    & $5.1 \pm 0.8$ \\
    IterativeAgent & $16.4 \pm 1.4$ \\
    PaperCoder \cite{seo2025paper2code}     & $44.2$ \\
    AutoP2C        & \textbf{$49.2 \pm 14.0$} \\  
    
    \bottomrule
    \end{tabular}%
    }
    \vspace{-0.1in}
\end{table}

\section{Experiments}
\subsection{Evaluation Metrics}

In our experiments, we evaluate the functional correctness and structural completeness of the code repositories generated by the AutoP2C framework.
For functional correctness, we execute the generated code repositories and the code of the original papers on the same datasets. 
To compare their performance, we introduce two traditional performance metrics, namely (1) absolute performance, which measures the outcomes (\textit{e}.\textit{g}., accuracy, F1 score, or other task-specific metrics) achieved by the AutoP2C generated code on designated datasets; and (2) relative performance, calculated as the ratio between the performance of the AutoP2C generated code and that of the original implementation. 
To evaluate structural completeness, \textit{i}.\textit{e}., how accurately the code structure captures the multimodal context in the academic paper, we define two new LLM-based metrics based on the LLM-as-a-Judge paradigm \cite{zheng2023judging, he2025code}, leveraging the LLM's outstanding performance in code quality evaluation.
The first is function completeness ($\text{COMP}_{\text{func}}$), which quantifies the quality of the implementation of the functions using a weighted LLM-based scoring, depicted as
\begin{equation}
\text{COMP}_{\text{func}} = \frac{\sum_{func_i \in Func} w_i \cdot \text{score}_{\text{func}}(f_i)}{\sum_{func_i \in Func} w_i},
\end{equation}
where $Func$ represents the set of functions in the code of the original paper, $w_i$ denotes the number of lines in function $func_i$, and $\text{score}_{\text{func}}(f_i) \in \{0, 0.2, 0.4, 0.6, 0.8, 1\}$ reflects implementation completeness on a 6-score scale given by LLM. 
The scoring ranges from unimplemented ($0$) to fully implemented ($1$), with intermediate values representing progressive levels of implementation quality: correct purpose but flawed logic ($0.2$), correct logic with major omissions ($0.4$), correct logic with partial omissions ($0.6$), and minor omissions ($0.8$).
Similarly, class completeness ($\text{COMP}_{\text{class}}$) evaluates how faithfully the generated code implements class structures specified in the paper's diagrams and text:
\begin{equation}
\text{COMP}_{\text{class}} = \frac{\sum_{c \in C} \sum_{m \in M_c} w_m \cdot \text{score}_{\text{method}}(m)}{\sum_{c \in C} \sum_{m \in M_c} w_m},
\end{equation}
where $C$ represents the set of classes, $M_c$ denotes the methods in class $c$, $w_m$ indicates the code lines of the method $m$, and $\text{score}_{\text{method}}(m)$ follows the same 6-point scoring system as $\text{score}_{\text{func}}$. 
Notably, if the implementations enhance functionality beyond the paper's specifications while maintaining alignment with its core principles, it will receive a $+0.2$ bonus, capped at $1.0$.
Additionally, we track computational resource utilization by measuring the number of input and output tokens throughout the entire process, which provides insights into the framework's efficiency when processing multimodal content with different complexities.


\subsection{Benchmark and Baseline}
The P2C task represents a novel challenge that extends beyond traditional code generation by requiring the integration of multimodal content into coherent, multi-file code repositories.
To provide a comprehensive evaluation, we consider two complementary benchmarks, Paper2Repo, and PaperBench Code-Dev \cite{starace2025paperbench}, each emphasizing a different aspect of repository quality.
(1) To evaluate how faithfully AutoP2C implements diverse models and algorithmic designs described in each paper, we constructed a brand-new benchmark called \textbf{Paper2Repo}, which comprises eight recent research papers all published after 2024 \cite{fontana2024distilled, nasiri2025edora, nazari2024entropy, xu2024no, liang20242gc, doshi2024reffakd, azimi2024kd, linse2023convolutional}. Each paper contains all the content required for the reproduction of the code repository.
These papers are selected from \url{paperswithcode.com}, and each has a corresponding GitHub repository as the upper bound for comparison. 
The benchmark spans six ML tasks (including training strategy optimization \cite{xu2024no}, node classification \cite{nazari2024entropy, liang20242gc}, model compression \cite{doshi2024reffakd}, parameter-efficient fine-tuning \cite{azimi2024kd, nasiri2025edora}, network pruning \cite{fontana2024distilled}, and image classification \cite{linse2023convolutional}) and three modal types of datasets (containing computer vision \cite{fontana2024distilled, xu2024no, doshi2024reffakd, linse2023convolutional}, natural language processing \cite{nasiri2025edora, azimi2024kd}, and graph \cite{nazari2024entropy, liang20242gc} related datasets).
We also established baselines using two state-of-the-art LLMs: OpenAI's recently released o1 \cite{zhong2024evaluation} and DeepSeek-R1 \cite{guo2025deepseek}, both known for their advanced code generation capabilities.
(2) To further assess the completeness of the code repositories generated by AutoP2C, we evaluated our framework on OpenAI’s newly released \textbf{PaperBench Code-Dev} benchmark \cite{starace2025paperbench}. We compared AutoP2C with several established baselines introduced in \cite{starace2025paperbench}, as well as PaperCoder, a recent multi-agent framework expressly designed to create code repositories from research papers.

\subsection{Experiment Settings}
From the repository blueprint extraction stage to the hierarchical task decomposition stage, we employ GPT-4o due to its advanced multimodal understanding capability.
For the initial file implementation, where code generation from processed specifications is the primary focus, we utilize o1-mini, a model optimized for code generation tasks. 
Then, we use o1 in the validation process, which is good at detecting nuanced logical inconsistencies.
Iterative code refinement is handled by o3-mini, which excels at refinement tasks. 
To ensure reproducibility and consistency across experiments, we standardize the decoding temperature to 0 for all temperature-configurable LLMs.
All experiments are conducted on an Intel(R) Xeon(R) Platinum 8352V CPU and an NVIDIA A40 (48GB) GPU.

\vspace{-0.1in}
\renewcommand{\arraystretch}{1.75} 
\newcolumntype{P}[1]{>{\centering\arraybackslash}p{#1}}

\begin{table*}[t]
    \centering
    \large
    \caption{Ablation study demonstrating the impact of each AutoP2C component on code generation performance.
    It examines four components, namely, repository blueprint extraction (blueprint), multimodal content parsing (MM), hierarchical task decomposition (Decop.), and iterative feedback-driven implementation (Feedback).
    Each row shows results when removing one component (marked with `x') while keeping others (marked with checkmarks). 
    The bottom row shows the complete AutoP2C framework with all components enabled. 
    Performance metrics include absolute accuracy (Perf.), relative performance compared to original implementations (Rel. Perf.), and structural fidelity metrics ($\text{COMP}_{\text{class}}$ and $\text{COMP}_{\text{func}}$). 
    Red crosses (\textcolor{red}{$\times$}) indicate failure to generate executable code. 
    \vspace{-0.1in}
    }
    \label{tab:ablation_1}
    \small
    \resizebox{\textwidth}{!}{%
    \begin{tabular}{%
  P{1.2cm} P{1.2cm} P{1.2cm} P{1.2cm}
  @{\hspace{6pt}}
  P{1.cm}@{\hspace{0pt}}P{1.6cm}@{\hspace{0pt}}P{1.6cm}@{\hspace{0pt}}P{1.6cm}
  @{\hspace{4pt}}
  P{1.cm}@{\hspace{0pt}}P{1.6cm}@{\hspace{0pt}}P{1.6cm}@{\hspace{0pt}}P{1.6cm}
  @{\hspace{4pt}}
  P{1.cm}@{\hspace{0pt}}P{1.6cm}@{\hspace{0pt}}P{1.6cm}@{\hspace{0pt}}P{1.6cm}
  @{\hspace{4pt}}
  P{1.cm}@{\hspace{0pt}}P{1.6cm}@{\hspace{0pt}}P{1.6cm}@{\hspace{0pt}}P{1.6cm}
    }
    \toprule
    \multicolumn{4}{c}{\textbf{Modules}} 
    & \multicolumn{4}{c}{\textbf{Paper's Title: Distilled... \cite{fontana2024distilled}}} 
    & \multicolumn{4}{c}{\textbf{Paper's Title: ReffAKD:... \cite{doshi2024reffakd}}} 
    & \multicolumn{4}{c}{\textbf{Paper's Title: $L^2$GC:... \cite{liang20242gc}}}
    & \multicolumn{4}{c}{\textbf{Paper's Title: Entropy Aware... \cite{nazari2024entropy}}} \\
    \cmidrule(lr){1-4} 
    \cmidrule(lr){5-8}
    \cmidrule(lr){9-12}
    \cmidrule(lr){13-16}
    \cmidrule(lr){17-20}
    \textbf{Blueprint} 
    & \textbf{MM} 
    & \textbf{Decomp.} 
    & \textbf{Feedback}
    & \textbf{Perf.} & \textbf{Rel. Perf.} & \textbf{\(\text{COMP}_{\text{class}}\)} & \textbf{\(\text{COMP}_{\text{func}}\)}
    & \textbf{Perf.} & \textbf{Rel. Perf.} & \textbf{\(\text{COMP}_{\text{class}}\)} & \textbf{\(\text{COMP}_{\text{func}}\)}
    & \textbf{Perf.} & \textbf{Rel. Perf.} & \textbf{\(\text{COMP}_{\text{class}}\)} & \textbf{\(\text{COMP}_{\text{func}}\)}
    & \textbf{Perf.} & \textbf{Rel. Perf.} & \textbf{\(\text{COMP}_{\text{class}}\)} & \textbf{\(\text{COMP}_{\text{func}}\)} \\
    \midrule
    x & \checkmark & \checkmark & \checkmark 
    & 10.0\% & 10.8\% & 90.8\% & 25.0\% 
    & 90.3\% & 96.6\% & 16.3\% & 32.0\%
    & 66.9\% & 81.2\% & 16.9\% & 18.1\%
    & 38.4\% & 49.5\% & 4.2\% & 17.9\% \\
    
    \checkmark & x & \checkmark & \checkmark 
    & 10.2\% & 11.0\% & 90.8\% & 44.1\% 
    & 83.5\% & 89.4\% & 16.0\% & 35.9\%
    & 70.2\% & 85.2\% & 0.0\% & 19.3\%
    & 50.1\% & 64.5\% & 45.1\% & 46.2\% \\
    
    \checkmark & \checkmark & x & \checkmark 
    & 66.2\% & 71.8\% & 90.8\% & 23.5\% 
    & 10.2\% & 10.9\% & 8.4\% & 32.5\%
    & 13.0\% & 15.8\% & 7.4\% & 9.6\%
    & 14.0\% & 18.0\% & 16.6\% & 13.6\% \\
    
    \checkmark & \checkmark & \checkmark & x 
    & \textcolor{red}{\(\times\)} & \textcolor{red}{\(\times\)} & 13.7\% & 24.6\% 
    & \textcolor{red}{\(\times\)} & \textcolor{red}{\(\times\)} & 16.1\% & 24.3\%
    & \textcolor{red}{\(\times\)} & \textcolor{red}{\(\times\)} & 0.0\% & 16.7\%
    & \textcolor{red}{\(\times\)} & \textcolor{red}{\(\times\)} & 33.6\% & 34.9\% \\
    
    \checkmark & \checkmark & \checkmark & \checkmark 
    & \textbf{82.8\%} & \textbf{89.8\%} & \textbf{90.8\%} & \textbf{60.2\%} 
    & \textbf{91.6\%} & \textbf{98.0\%} & \textbf{25.9\%} & \textbf{35.9\%}
    & \textbf{79.5\%} & \textbf{96.5\%} & \textbf{75.9\%} & \textbf{31.4\%}
    & \textbf{79.9\%} & \textbf{103.0\%} & \textbf{53.8\%} & \textbf{46.2\%} \\
    \bottomrule
    \end{tabular}%
    }
    \vspace{-0.1in}
\end{table*}
\subsection{Results}
\subsubsection{Evaluation for Functional Correctness}
The results in Table~\ref{tab:merged} demonstrate the advantages of AutoP2C in both implementation feasibility and functional fidelity compared to the baselines. 
While OpenAI-o1 and DeepSeek-R1 successfully generate executable code for only one paper (``No More Adam: ...'' \cite{xu2024no} ), AutoP2C produces functional implementations across all eight research papers in our benchmark. 
This contrast in success rate (100\% versus 12.5\%) underscores the effectiveness of AutoP2C in performing the P2C task.
Even in the single case where the baseline models produce executable code, AutoP2C achieves superior performance. 
Specifically, for ``No More Adam: ...'' \cite{xu2024no}, our AutoP2C delivers 91.4\% accuracy compared to 81.8\% for o1 (9.6\% improvement) and 75.3\% for R1 (16.1\% improvement). 
The relative performance metric further validates the effectiveness of our framework, with AutoP2C achieving between 89.8\% and 122.0\% of the performance of the original implementations in all eight papers. 
Notably, in two cases, the performance of the code repositories generated by AutoP2C exceeds that of the code of the original papers, namely ``Entropy Aware..'' \cite{nazari2024entropy} (103.0\%) and ``Convolutional...'' \cite{linse2023convolutional} (122.0\%). 
The consistently high relative performance (averaging 99.5\% across all papers) confirms that AutoP2C's staged generation paradigm ensures both compliance with paper specifications and operational integrity in real-world execution. 
Additionally, we provide a case study of AutoP2C in Fig. \ref{fig:wide_example2}, demonstrating how our approach successfully transforms multimodal research content into executable code through its four-stage pipeline.

\subsubsection{Evaluation for Structural Completeness}  

The structural completeness metrics in Table~\ref{tab:merged} provide quantitative evidence of AutoP2C's superior ability. 
For class completeness ($\text{COMP}_{\text{class}}$), AutoP2C consistently outperforms both baseline models across all papers where class implementations were relevant. 
Our framework achieves an average score of 65.7\%, representing a $1.9\times$ increase compared to OpenAI-o1 (34.9\%) and a $2.1\times$ increase compared to DeepSeek-R1 (31.1\%), which demonstrates the effectiveness of AutoP2C in correctly interpreting architectural diagrams and translating them into appropriate class hierarchies and relationships.
Similarly, for function completeness ($\text{COMP}_{\text{func}}$), AutoP2C delivers superior results in all applicable cases. 
On average, AutoP2C achieves a score of 51.6\%, surpassing o1 (29.1\%) by $1.8\times$ and R1 (17.6\%) by $2.9\times$. 
The performance gap between AutoP2C and the baselines becomes even more pronounced when handling papers with complex mathematical notation. 
For graph-related papers specifically \cite{nazari2024entropy, liang20242gc}, which typically contain extensive mathematical formulations, AutoP2C's $\text{COMP}_{\text{class}}$ score is $4.9\times$ higher than o1 and $2.4\times$ higher than R1. 
Similarly, its $\text{COMP}_{\text{func}}$ score for these papers is $2.8\times$ higher than o1 and $3.8\times$ higher than R1. 
These substantial differences highlight AutoP2C's particular strength in processing mathematically intensive content that requires careful integration of multiple modalities.

\subsubsection{Evaluation for PaperBench Code-Dev}
To gauge how well AutoP2C generalizes beyond our own curated corpus, we further benchmark it on PaperBench Code-Dev, a public suite that automatically grades end-to-end reproduction of research code.
As summarised in Table ~\ref{tab:code_dev_scores}, AutoP2C attains the highest average replication score, $49.2 \pm 14.0\%$, surpassing PaperCoder (44.2\%) by roughly five percentage points and outperforming the lightweight BasicAgent ($5.1 \pm 0.8\%$) and IterativeAgent ($16.4 \pm 1.4 \%$) by wide margins.
These results confirm that our structured multi-agent pipeline translates more of a paper’s specification into functional, runnable code than existing single- or few-agent baselines.

\renewcommand{\arraystretch}{1.75} 
\setlength{\tabcolsep}{0.5pt}     
\newcolumntype{P}[1]{>{\centering\arraybackslash}p{#1}}

\begin{table}[t]
    \centering
    \large
    \caption{Ablation study on the contribution of different modalities in multimodal content phrasing stage, including architectural diagrams (image) and experimental results (table) when processing the ``Convolutional...'' \cite{liang20242gc} paper. 
    Checkmarks indicate included modalities, while `x' denotes excluded ones. 
    }
    \vspace{-0.1in}
    \label{tab:ablation_2}
    \small
    \resizebox{\columnwidth}{!}{%
    \begin{tabular}{%
    P{1.2cm} P{1.2cm} P{1.2cm}
    @{\hspace{6pt}}
    P{1.0cm}@{\hspace{0pt}}P{1.6cm}@{\hspace{0pt}}P{1.6cm}@{\hspace{0pt}}P{1.6cm}
    }
    \toprule
    \multicolumn{3}{c}{\textbf{Modality}} 
    & \multicolumn{4}{c}{\textbf{Paper's Title: Convolutional...}} \\
    \cmidrule(lr){1-3}
    \cmidrule(lr){4-7}
    \textbf{Text} & \textbf{Image} & \textbf{Table}
    & \textbf{Perf.} & \textbf{Rel. Perf.} & \textbf{\(\text{COMP}_{\text{class}}\)} & \textbf{\(\text{COMP}_{\text{func}}\)} \\
    \midrule
    
    \checkmark & x & \checkmark
    & 70.1\% & 92.9\% & 59.8\% & 18.5\% \\
    
    \checkmark & \checkmark & x
    & 88.9\% & 117.9\% & 36.7\% & 7.6\% \\
    
    \checkmark & \checkmark & \checkmark
    & \textbf{92.0\%} & \textbf{122.0\%} & \textbf{73.1\%} & \textbf{18.5\%} \\
    
    \bottomrule
    \end{tabular}%
    }
    \vspace{-0.1in}
\end{table}

\subsection{Ablation Study}

The ablation analysis in Table \ref{tab:ablation_1} reveals that the removal of any component degrades performance.
When the repository blueprint extraction component is removed, the LLM must construct file organization structures without guidance, resulting in disorganized and redundant code. 
This absence of a structured template causes average performance to decline from 83.5\% to 51.4\%. 
The structural completeness metrics also decline, with the average class completeness dropping from 61.6\% to 32.1\% and the completeness of functions decreasing from 43.4\% to 23.2\%. 
The multimodal content parsing component is equally important. 
In particular, the absence of visual understanding and redundancy removing results in average performance declining to 53.5\%, with class completeness and function completeness falling to 38.0\% and 36.4\% respectively.
Eliminating the hierarchical task decomposition phase forces the model to generate entire implementations in a single pass, which results in the average class completeness decreasing to 30.8\% and produces the lowest average function completeness (19.8\%) scores across our ablation. 
Without the iterative feedback-driven implementation, AutoP2C relies entirely on single-pass generation, resulting in completely non-executable implementations across all test cases and yielding the lowest average class completeness of 15.9\%. Moreover, the function completeness falls to 25.1\%.

Table \ref{tab:ablation_2} presents another ablation study on the contribution of different modalities in the multimodal content parsing stage. 
The results demonstrate that each modality provides important information for generating accurate and complete code repositories.
When architectural diagrams (image modality) are excluded from the input, performance decreases from 92.0\% to 70.1\% (a 21.9\% absolute reduction), while class completeness declines from 73.1\% to 59.8\%. 
This substantial degradation occurs because architectural diagrams contain essential visual information about model structure, component relationships, and data flow that cannot be fully captured by textual descriptions alone. 
The absence of tabular modality produces different but equally important effects.
These findings confirm the need for multimodal understanding in the P2C task. 

\section{Conclusion}
This paper introduces ``Paper-to-Code'' (P2C), a novel task that addresses the challenging problem of translating multimodal research content from academic papers into fully executable code repositories. 
To tackle this task, we propose AutoP2C, a multi-agent framework that employs LLMs, which represents an advancement toward improving research reproducibility. 
Future work could explore the extension of AutoP2C to support additional programming languages beyond Python and incorporate user feedback to continuously improve the generated code repositories.

\bibliographystyle{ACM-Reference-Format}
\bibliography{sample-base}


\begin{thebibliography}{68}


\ifx \showCODEN    \undefined \def \showCODEN     #1{\unskip}     \fi
\ifx \showDOI      \undefined \def \showDOI       #1{#1}\fi
\ifx \showISBNx    \undefined \def \showISBNx     #1{\unskip}     \fi
\ifx \showISBNxiii \undefined \def \showISBNxiii  #1{\unskip}     \fi
\ifx \showISSN     \undefined \def \showISSN      #1{\unskip}     \fi
\ifx \showLCCN     \undefined \def \showLCCN      #1{\unskip}     \fi
\ifx \shownote     \undefined \def \shownote      #1{#1}          \fi
\ifx \showarticletitle \undefined \def \showarticletitle #1{#1}   \fi
\ifx \showURL      \undefined \def \showURL       {\relax}        \fi
\providecommand\bibfield[2]{#2}
\providecommand\bibinfo[2]{#2}
\providecommand\natexlab[1]{#1}
\providecommand\showeprint[2][]{arXiv:#2}

\bibitem[Ali and Smith(2006)]%
        {ali2006learning}
\bibfield{author}{\bibinfo{person}{Shawkat Ali} {and} \bibinfo{person}{Kate~A Smith}.} \bibinfo{year}{2006}\natexlab{}.
\newblock \showarticletitle{On learning algorithm selection for classification}.
\newblock \bibinfo{journal}{\emph{Applied Soft Computing}} \bibinfo{volume}{6}, \bibinfo{number}{2} (\bibinfo{year}{2006}), \bibinfo{pages}{119--138}.
\newblock


\bibitem[Anil et~al\mbox{.}(2022)]%
        {anil2022exploring}
\bibfield{author}{\bibinfo{person}{Cem Anil}, \bibinfo{person}{Yuhuai Wu}, \bibinfo{person}{Anders Andreassen}, \bibinfo{person}{Aitor Lewkowycz}, \bibinfo{person}{Vedant Misra}, \bibinfo{person}{Vinay Ramasesh}, \bibinfo{person}{Ambrose Slone}, \bibinfo{person}{Guy Gur-Ari}, \bibinfo{person}{Ethan Dyer}, {and} \bibinfo{person}{Behnam Neyshabur}.} \bibinfo{year}{2022}\natexlab{}.
\newblock \showarticletitle{Exploring length generalization in large language models}.
\newblock \bibinfo{journal}{\emph{Advances in Neural Information Processing Systems}}  \bibinfo{volume}{35} (\bibinfo{year}{2022}), \bibinfo{pages}{38546--38556}.
\newblock


\bibitem[Austin et~al\mbox{.}(2021)]%
        {austin2021program}
\bibfield{author}{\bibinfo{person}{Jacob Austin}, \bibinfo{person}{Augustus Odena}, \bibinfo{person}{Maxwell Nye}, \bibinfo{person}{Maarten Bosma}, \bibinfo{person}{Henryk Michalewski}, \bibinfo{person}{David Dohan}, \bibinfo{person}{Ellen Jiang}, \bibinfo{person}{Carrie Cai}, \bibinfo{person}{Michael Terry}, \bibinfo{person}{Quoc Le}, {et~al\mbox{.}}} \bibinfo{year}{2021}\natexlab{}.
\newblock \showarticletitle{Program synthesis with large language models}.
\newblock \bibinfo{journal}{\emph{arXiv preprint arXiv:2108.07732}} (\bibinfo{year}{2021}).
\newblock


\bibitem[Ayanwale et~al\mbox{.}(2024)]%
        {ayanwale2024analyzing}
\bibfield{author}{\bibinfo{person}{Musa~Adekunle Ayanwale}, \bibinfo{person}{Rethabile~Rosemary Molefi}, {and} \bibinfo{person}{Saheed Oyeniran}.} \bibinfo{year}{2024}\natexlab{}.
\newblock \showarticletitle{Analyzing the evolution of machine learning integration in educational research: a bibliometric perspective}.
\newblock \bibinfo{journal}{\emph{Discover Education}} \bibinfo{volume}{3}, \bibinfo{number}{1} (\bibinfo{year}{2024}), \bibinfo{pages}{47}.
\newblock


\bibitem[Azimi et~al\mbox{.}(2024)]%
        {azimi2024kd}
\bibfield{author}{\bibinfo{person}{Rambod Azimi}, \bibinfo{person}{Rishav Rishav}, \bibinfo{person}{Marek Teichmann}, {and} \bibinfo{person}{Samira~Ebrahimi Kahou}.} \bibinfo{year}{2024}\natexlab{}.
\newblock \showarticletitle{KD-LoRA: A Hybrid Approach to Efficient Fine-Tuning with LoRA and Knowledge Distillation}.
\newblock \bibinfo{journal}{\emph{arXiv preprint arXiv:2410.20777}} (\bibinfo{year}{2024}).
\newblock


\bibitem[Ball(2023)]%
        {ball2023ai}
\bibfield{author}{\bibinfo{person}{Philip Ball}.} \bibinfo{year}{2023}\natexlab{}.
\newblock \showarticletitle{Is AI leading to a reproducibility crisis in science?}
\newblock \bibinfo{journal}{\emph{Nature}} \bibinfo{volume}{624}, \bibinfo{number}{7990} (\bibinfo{year}{2023}), \bibinfo{pages}{22--25}.
\newblock


\bibitem[Bergstra and Bengio(2012)]%
        {bergstra2012random}
\bibfield{author}{\bibinfo{person}{James Bergstra} {and} \bibinfo{person}{Yoshua Bengio}.} \bibinfo{year}{2012}\natexlab{}.
\newblock \showarticletitle{Random search for hyper-parameter optimization}.
\newblock \bibinfo{journal}{\emph{The journal of machine learning research}} \bibinfo{volume}{13}, \bibinfo{number}{1} (\bibinfo{year}{2012}), \bibinfo{pages}{281--305}.
\newblock


\bibitem[Bisong and Bisong(2019)]%
        {bisong2019google}
\bibfield{author}{\bibinfo{person}{Ekaba Bisong} {and} \bibinfo{person}{Ekaba Bisong}.} \bibinfo{year}{2019}\natexlab{}.
\newblock \showarticletitle{Google automl: cloud vision}.
\newblock \bibinfo{journal}{\emph{Building Machine Learning and Deep Learning Models on Google Cloud Platform: A Comprehensive Guide for Beginners}} (\bibinfo{year}{2019}), \bibinfo{pages}{581--598}.
\newblock


\bibitem[Blecher et~al\mbox{.}(2023)]%
        {blecher2023nougat}
\bibfield{author}{\bibinfo{person}{Lukas Blecher}, \bibinfo{person}{Guillem Cucurull}, \bibinfo{person}{Thomas Scialom}, {and} \bibinfo{person}{Robert Stojnic}.} \bibinfo{year}{2023}\natexlab{}.
\newblock \showarticletitle{Nougat: Neural optical understanding for academic documents}.
\newblock \bibinfo{journal}{\emph{arXiv preprint arXiv:2308.13418}} (\bibinfo{year}{2023}).
\newblock


\bibitem[Chen et~al\mbox{.}(2023)]%
        {chen2023autoagents}
\bibfield{author}{\bibinfo{person}{Guangyao Chen}, \bibinfo{person}{Siwei Dong}, \bibinfo{person}{Yu Shu}, \bibinfo{person}{Ge Zhang}, \bibinfo{person}{Jaward Sesay}, \bibinfo{person}{B{\"o}rje~F Karlsson}, \bibinfo{person}{Jie Fu}, {and} \bibinfo{person}{Yemin Shi}.} \bibinfo{year}{2023}\natexlab{}.
\newblock \showarticletitle{Autoagents: A framework for automatic agent generation}.
\newblock \bibinfo{journal}{\emph{arXiv preprint arXiv:2309.17288}} (\bibinfo{year}{2023}).
\newblock


\bibitem[Chen et~al\mbox{.}(2025)]%
        {chen2025ocean}
\bibfield{author}{\bibinfo{person}{Song Chen}, \bibinfo{person}{Xinyu Guo}, \bibinfo{person}{Yadong Li}, \bibinfo{person}{Tao Zhang}, \bibinfo{person}{Mingan Lin}, \bibinfo{person}{Dongdong Kuang}, \bibinfo{person}{Youwei Zhang}, \bibinfo{person}{Lingfeng Ming}, \bibinfo{person}{Fengyu Zhang}, \bibinfo{person}{Yuran Wang}, {et~al\mbox{.}}} \bibinfo{year}{2025}\natexlab{}.
\newblock \showarticletitle{Ocean-OCR: Towards General OCR Application via a Vision-Language Model}.
\newblock \bibinfo{journal}{\emph{arXiv preprint arXiv:2501.15558}} (\bibinfo{year}{2025}).
\newblock


\bibitem[Chu et~al\mbox{.}(2024)]%
        {chu2024mobilevlm}
\bibfield{author}{\bibinfo{person}{Xiangxiang Chu}, \bibinfo{person}{Limeng Qiao}, \bibinfo{person}{Xinyu Zhang}, \bibinfo{person}{Shuang Xu}, \bibinfo{person}{Fei Wei}, \bibinfo{person}{Yang Yang}, \bibinfo{person}{Xiaofei Sun}, \bibinfo{person}{Yiming Hu}, \bibinfo{person}{Xinyang Lin}, \bibinfo{person}{Bo Zhang}, {et~al\mbox{.}}} \bibinfo{year}{2024}\natexlab{}.
\newblock \showarticletitle{Mobilevlm v2: Faster and stronger baseline for vision language model}.
\newblock \bibinfo{journal}{\emph{arXiv preprint arXiv:2402.03766}} (\bibinfo{year}{2024}).
\newblock


\bibitem[Das et~al\mbox{.}(2020)]%
        {das2020amazon}
\bibfield{author}{\bibinfo{person}{Piali Das}, \bibinfo{person}{Nikita Ivkin}, \bibinfo{person}{Tanya Bansal}, \bibinfo{person}{Laurence Rouesnel}, \bibinfo{person}{Philip Gautier}, \bibinfo{person}{Zohar Karnin}, \bibinfo{person}{Leo Dirac}, \bibinfo{person}{Lakshmi Ramakrishnan}, \bibinfo{person}{Andre Perunicic}, \bibinfo{person}{Iaroslav Shcherbatyi}, {et~al\mbox{.}}} \bibinfo{year}{2020}\natexlab{}.
\newblock \showarticletitle{Amazon SageMaker Autopilot: a white box AutoML solution at scale}. In \bibinfo{booktitle}{\emph{Proceedings of the fourth international workshop on data management for end-to-end machine learning}}. \bibinfo{pages}{1--7}.
\newblock


\bibitem[Dong et~al\mbox{.}(2024)]%
        {dong2024self}
\bibfield{author}{\bibinfo{person}{Yihong Dong}, \bibinfo{person}{Xue Jiang}, \bibinfo{person}{Zhi Jin}, {and} \bibinfo{person}{Ge Li}.} \bibinfo{year}{2024}\natexlab{}.
\newblock \showarticletitle{Self-collaboration code generation via chatgpt}.
\newblock \bibinfo{journal}{\emph{ACM Transactions on Software Engineering and Methodology}} \bibinfo{volume}{33}, \bibinfo{number}{7} (\bibinfo{year}{2024}), \bibinfo{pages}{1--38}.
\newblock


\bibitem[Doshi and Kim(2024)]%
        {doshi2024reffakd}
\bibfield{author}{\bibinfo{person}{Divyang Doshi} {and} \bibinfo{person}{Jung-Eun Kim}.} \bibinfo{year}{2024}\natexlab{}.
\newblock \showarticletitle{ReffAKD: Resource-efficient Autoencoder-based Knowledge Distillation}.
\newblock \bibinfo{journal}{\emph{arXiv preprint arXiv:2404.09886}} (\bibinfo{year}{2024}).
\newblock


\bibitem[Fontana et~al\mbox{.}(2024)]%
        {fontana2024distilled}
\bibfield{author}{\bibinfo{person}{Federico Fontana}, \bibinfo{person}{Romeo Lanzino}, \bibinfo{person}{Marco~Raoul Marini}, \bibinfo{person}{Danilo Avola}, \bibinfo{person}{Luigi Cinque}, \bibinfo{person}{Francesco Scarcello}, {and} \bibinfo{person}{Gian~Luca Foresti}.} \bibinfo{year}{2024}\natexlab{}.
\newblock \showarticletitle{Distilled gradual pruning with pruned fine-tuning}.
\newblock \bibinfo{journal}{\emph{IEEE Transactions on Artificial Intelligence}} (\bibinfo{year}{2024}).
\newblock


\bibitem[Gong et~al\mbox{.}(2024)]%
        {gong2024evaluation}
\bibfield{author}{\bibinfo{person}{Linyuan Gong}, \bibinfo{person}{Sida Wang}, \bibinfo{person}{Mostafa Elhoushi}, {and} \bibinfo{person}{Alvin Cheung}.} \bibinfo{year}{2024}\natexlab{}.
\newblock \showarticletitle{Evaluation of llms on syntax-aware code fill-in-the-middle tasks}.
\newblock \bibinfo{journal}{\emph{arXiv preprint arXiv:2403.04814}} (\bibinfo{year}{2024}).
\newblock


\bibitem[Guo et~al\mbox{.}(2025)]%
        {guo2025deepseek}
\bibfield{author}{\bibinfo{person}{Daya Guo}, \bibinfo{person}{Dejian Yang}, \bibinfo{person}{Haowei Zhang}, \bibinfo{person}{Junxiao Song}, \bibinfo{person}{Ruoyu Zhang}, \bibinfo{person}{Runxin Xu}, \bibinfo{person}{Qihao Zhu}, \bibinfo{person}{Shirong Ma}, \bibinfo{person}{Peiyi Wang}, \bibinfo{person}{Xiao Bi}, {et~al\mbox{.}}} \bibinfo{year}{2025}\natexlab{}.
\newblock \showarticletitle{Deepseek-r1: Incentivizing reasoning capability in llms via reinforcement learning}.
\newblock \bibinfo{journal}{\emph{arXiv preprint arXiv:2501.12948}} (\bibinfo{year}{2025}).
\newblock


\bibitem[Guo et~al\mbox{.}(2024)]%
        {guo2024deepseek}
\bibfield{author}{\bibinfo{person}{Daya Guo}, \bibinfo{person}{Qihao Zhu}, \bibinfo{person}{Dejian Yang}, \bibinfo{person}{Zhenda Xie}, \bibinfo{person}{Kai Dong}, \bibinfo{person}{Wentao Zhang}, \bibinfo{person}{Guanting Chen}, \bibinfo{person}{Xiao Bi}, \bibinfo{person}{Yu Wu}, \bibinfo{person}{YK Li}, {et~al\mbox{.}}} \bibinfo{year}{2024}\natexlab{}.
\newblock \showarticletitle{DeepSeek-Coder: When the Large Language Model Meets Programming--The Rise of Code Intelligence}.
\newblock \bibinfo{journal}{\emph{arXiv preprint arXiv:2401.14196}} (\bibinfo{year}{2024}).
\newblock


\bibitem[Hanson et~al\mbox{.}(2024)]%
        {hanson2024strain}
\bibfield{author}{\bibinfo{person}{Mark~A Hanson}, \bibinfo{person}{Pablo~G{\'o}mez Barreiro}, \bibinfo{person}{Paolo Crosetto}, {and} \bibinfo{person}{Dan Brockington}.} \bibinfo{year}{2024}\natexlab{}.
\newblock \showarticletitle{The strain on scientific publishing}.
\newblock \bibinfo{journal}{\emph{Quantitative Science Studies}} \bibinfo{volume}{5}, \bibinfo{number}{4} (\bibinfo{year}{2024}), \bibinfo{pages}{823--843}.
\newblock


\bibitem[He et~al\mbox{.}(2025)]%
        {he2025code}
\bibfield{author}{\bibinfo{person}{Junda He}, \bibinfo{person}{Jieke Shi}, \bibinfo{person}{Terry~Yue Zhuo}, \bibinfo{person}{Christoph Treude}, \bibinfo{person}{Jiamou Sun}, \bibinfo{person}{Zhenchang Xing}, \bibinfo{person}{Xiaoning Du}, {and} \bibinfo{person}{David Lo}.} \bibinfo{year}{2025}\natexlab{}.
\newblock \showarticletitle{From code to courtroom: Llms as the new software judges}.
\newblock \bibinfo{journal}{\emph{arXiv preprint arXiv:2503.02246}} (\bibinfo{year}{2025}).
\newblock


\bibitem[Hong et~al\mbox{.}(2024)]%
        {Hong2024}
\bibfield{author}{\bibinfo{person}{Sirui Hong}, \bibinfo{person}{Mingchen Zhuge}, \bibinfo{person}{Jonathan Chen}, {et~al\mbox{.}}} \bibinfo{year}{2024}\natexlab{}.
\newblock \showarticletitle{MetaGPT: Meta Programming for A Multi-Agent Collaborative Framework}. In \bibinfo{booktitle}{\emph{Proceedings of ICLR 2024}}.
\newblock


\bibitem[Huang et~al\mbox{.}(2023)]%
        {huang2023agentcoder}
\bibfield{author}{\bibinfo{person}{Dong Huang}, \bibinfo{person}{Jie~M Zhang}, \bibinfo{person}{Michael Luck}, \bibinfo{person}{Qingwen Bu}, \bibinfo{person}{Yuhao Qing}, {and} \bibinfo{person}{Heming Cui}.} \bibinfo{year}{2023}\natexlab{}.
\newblock \showarticletitle{Agentcoder: Multi-agent-based code generation with iterative testing and optimisation}.
\newblock \bibinfo{journal}{\emph{arXiv preprint arXiv:2312.13010}} (\bibinfo{year}{2023}).
\newblock


\bibitem[Ishibashi and Nishimura(2024)]%
        {ishibashi2024self}
\bibfield{author}{\bibinfo{person}{Yoichi Ishibashi} {and} \bibinfo{person}{Yoshimasa Nishimura}.} \bibinfo{year}{2024}\natexlab{}.
\newblock \showarticletitle{Self-organized agents: A llm multi-agent framework toward ultra large-scale code generation and optimization}.
\newblock \bibinfo{journal}{\emph{arXiv preprint arXiv:2404.02183}} (\bibinfo{year}{2024}).
\newblock


\bibitem[Islam et~al\mbox{.}(2024)]%
        {islam2024mapcoder}
\bibfield{author}{\bibinfo{person}{Md~Ashraful Islam}, \bibinfo{person}{Mohammed~Eunus Ali}, {and} \bibinfo{person}{Md~Rizwan Parvez}.} \bibinfo{year}{2024}\natexlab{}.
\newblock \showarticletitle{Mapcoder: Multi-agent code generation for competitive problem solving}.
\newblock \bibinfo{journal}{\emph{arXiv preprint arXiv:2405.11403}} (\bibinfo{year}{2024}).
\newblock


\bibitem[Ji et~al\mbox{.}(2023)]%
        {ji2023survey}
\bibfield{author}{\bibinfo{person}{Ziwei Ji}, \bibinfo{person}{Nayeon Lee}, \bibinfo{person}{Rita Frieske}, \bibinfo{person}{Tiezheng Yu}, \bibinfo{person}{Dan Su}, \bibinfo{person}{Yan Xu}, \bibinfo{person}{Etsuko Ishii}, \bibinfo{person}{Ye~Jin Bang}, \bibinfo{person}{Andrea Madotto}, {and} \bibinfo{person}{Pascale Fung}.} \bibinfo{year}{2023}\natexlab{}.
\newblock \showarticletitle{Survey of hallucination in natural language generation}.
\newblock \bibinfo{journal}{\emph{ACM computing surveys}} \bibinfo{volume}{55}, \bibinfo{number}{12} (\bibinfo{year}{2023}), \bibinfo{pages}{1--38}.
\newblock


\bibitem[Jiang et~al\mbox{.}(2024)]%
        {jiang2024self}
\bibfield{author}{\bibinfo{person}{Xue Jiang}, \bibinfo{person}{Yihong Dong}, \bibinfo{person}{Lecheng Wang}, \bibinfo{person}{Zheng Fang}, \bibinfo{person}{Qiwei Shang}, \bibinfo{person}{Ge Li}, \bibinfo{person}{Zhi Jin}, {and} \bibinfo{person}{Wenpin Jiao}.} \bibinfo{year}{2024}\natexlab{}.
\newblock \showarticletitle{Self-planning code generation with large language models}.
\newblock \bibinfo{journal}{\emph{ACM Transactions on Software Engineering and Methodology}} \bibinfo{volume}{33}, \bibinfo{number}{7} (\bibinfo{year}{2024}), \bibinfo{pages}{1--30}.
\newblock


\bibitem[Jin et~al\mbox{.}(2019)]%
        {jin2019auto}
\bibfield{author}{\bibinfo{person}{Haifeng Jin}, \bibinfo{person}{Qingquan Song}, {and} \bibinfo{person}{Xia Hu}.} \bibinfo{year}{2019}\natexlab{}.
\newblock \showarticletitle{Auto-keras: An efficient neural architecture search system}. In \bibinfo{booktitle}{\emph{Proceedings of the 25th ACM SIGKDD international conference on knowledge discovery \& data mining}}. \bibinfo{pages}{1946--1956}.
\newblock


\bibitem[LaBash et~al\mbox{.}(2024)]%
        {labash2024res}
\bibfield{author}{\bibinfo{person}{Beck LaBash}, \bibinfo{person}{August Rosedale}, \bibinfo{person}{Alex Reents}, \bibinfo{person}{Lucas Negritto}, {and} \bibinfo{person}{Colin Wiel}.} \bibinfo{year}{2024}\natexlab{}.
\newblock \showarticletitle{RES-Q: Evaluating Code-Editing Large Language Model Systems at the Repository Scale}.
\newblock \bibinfo{journal}{\emph{arXiv preprint arXiv:2406.16801}} (\bibinfo{year}{2024}).
\newblock


\bibitem[Le et~al\mbox{.}(2022)]%
        {le2022coderl}
\bibfield{author}{\bibinfo{person}{Hung Le}, \bibinfo{person}{Yue Wang}, \bibinfo{person}{Akhilesh~Deepak Gotmare}, \bibinfo{person}{Silvio Savarese}, {and} \bibinfo{person}{Steven Chu~Hong Hoi}.} \bibinfo{year}{2022}\natexlab{}.
\newblock \showarticletitle{Coderl: Mastering code generation through pretrained models and deep reinforcement learning}.
\newblock \bibinfo{journal}{\emph{Advances in Neural Information Processing Systems}}  \bibinfo{volume}{35} (\bibinfo{year}{2022}), \bibinfo{pages}{21314--21328}.
\newblock


\bibitem[Li et~al\mbox{.}(2024)]%
        {li2024mmcode}
\bibfield{author}{\bibinfo{person}{Kaixin Li}, \bibinfo{person}{Yuchen Tian}, \bibinfo{person}{Qisheng Hu}, \bibinfo{person}{Ziyang Luo}, \bibinfo{person}{Zhiyong Huang}, {and} \bibinfo{person}{Jing Ma}.} \bibinfo{year}{2024}\natexlab{}.
\newblock \showarticletitle{MMCode: Benchmarking Multimodal Large Language Models for Code Generation with Visually Rich Programming Problems}.
\newblock \bibinfo{journal}{\emph{arXiv preprint arXiv:2404.09486}} (\bibinfo{year}{2024}).
\newblock


\bibitem[Li et~al\mbox{.}(2022)]%
        {li2022identifying}
\bibfield{author}{\bibinfo{person}{Xin Li}, \bibinfo{person}{Yang Wen}, \bibinfo{person}{Jiaojiao Jiang}, \bibinfo{person}{Tugrul Daim}, {and} \bibinfo{person}{Lucheng Huang}.} \bibinfo{year}{2022}\natexlab{}.
\newblock \showarticletitle{Identifying potential breakthrough research: A machine learning method using scientific papers and Twitter data}.
\newblock \bibinfo{journal}{\emph{Technological Forecasting and Social Change}}  \bibinfo{volume}{184} (\bibinfo{year}{2022}), \bibinfo{pages}{122042}.
\newblock


\bibitem[Liang et~al\mbox{.}(2024)]%
        {liang20242gc}
\bibfield{author}{\bibinfo{person}{Qiuyu Liang}, \bibinfo{person}{Weihua Wang}, \bibinfo{person}{Feilong Bao}, {and} \bibinfo{person}{Guanglai Gao}.} \bibinfo{year}{2024}\natexlab{}.
\newblock \showarticletitle{L\^{} 2GC: Lorentzian linear graph convolutional networks for node classification}.
\newblock \bibinfo{journal}{\emph{arXiv preprint arXiv:2403.06064}} (\bibinfo{year}{2024}).
\newblock


\bibitem[Linse et~al\mbox{.}(2023)]%
        {linse2023convolutional}
\bibfield{author}{\bibinfo{person}{Christoph Linse}, \bibinfo{person}{Erhardt Barth}, {and} \bibinfo{person}{Thomas Martinetz}.} \bibinfo{year}{2023}\natexlab{}.
\newblock \showarticletitle{Convolutional neural networks do work with pre-defined filters}. In \bibinfo{booktitle}{\emph{2023 International Joint Conference on Neural Networks (IJCNN)}}. IEEE, \bibinfo{pages}{1--8}.
\newblock


\bibitem[Liu et~al\mbox{.}(2023)]%
        {liu2023your}
\bibfield{author}{\bibinfo{person}{Jiawei Liu}, \bibinfo{person}{Chunqiu~Steven Xia}, \bibinfo{person}{Yuyao Wang}, {and} \bibinfo{person}{Lingming Zhang}.} \bibinfo{year}{2023}\natexlab{}.
\newblock \showarticletitle{Is your code generated by chatgpt really correct? rigorous evaluation of large language models for code generation}.
\newblock \bibinfo{journal}{\emph{Advances in Neural Information Processing Systems}}  \bibinfo{volume}{36} (\bibinfo{year}{2023}), \bibinfo{pages}{21558--21572}.
\newblock


\bibitem[Liu et~al\mbox{.}(2024)]%
        {liu2024ocrbench}
\bibfield{author}{\bibinfo{person}{Yuliang Liu}, \bibinfo{person}{Zhang Li}, \bibinfo{person}{Mingxin Huang}, \bibinfo{person}{Biao Yang}, \bibinfo{person}{Wenwen Yu}, \bibinfo{person}{Chunyuan Li}, \bibinfo{person}{Xu-Cheng Yin}, \bibinfo{person}{Cheng-Lin Liu}, \bibinfo{person}{Lianwen Jin}, {and} \bibinfo{person}{Xiang Bai}.} \bibinfo{year}{2024}\natexlab{}.
\newblock \showarticletitle{OCRBench: on the hidden mystery of OCR in large multimodal models}.
\newblock \bibinfo{journal}{\emph{Science China Information Sciences}} \bibinfo{volume}{67}, \bibinfo{number}{12} (\bibinfo{year}{2024}), \bibinfo{pages}{220102}.
\newblock


\bibitem[Lu et~al\mbox{.}(2021)]%
        {lu2021codexglue}
\bibfield{author}{\bibinfo{person}{Shuai Lu}, \bibinfo{person}{Daya Guo}, \bibinfo{person}{Shuo Ren}, \bibinfo{person}{Junjie Huang}, \bibinfo{person}{Alexey Svyatkovskiy}, \bibinfo{person}{Ambrosio Blanco}, \bibinfo{person}{Colin Clement}, \bibinfo{person}{Dawn Drain}, \bibinfo{person}{Daxin Jiang}, \bibinfo{person}{Duyu Tang}, {et~al\mbox{.}}} \bibinfo{year}{2021}\natexlab{}.
\newblock \showarticletitle{Codexglue: A machine learning benchmark dataset for code understanding and generation}.
\newblock \bibinfo{journal}{\emph{arXiv preprint arXiv:2102.04664}} (\bibinfo{year}{2021}).
\newblock


\bibitem[Luo et~al\mbox{.}(2024)]%
        {luo2024autom3l}
\bibfield{author}{\bibinfo{person}{Daqin Luo}, \bibinfo{person}{Chengjian Feng}, \bibinfo{person}{Yuxuan Nong}, {and} \bibinfo{person}{Yiqing Shen}.} \bibinfo{year}{2024}\natexlab{}.
\newblock \showarticletitle{Autom3l: An automated multimodal machine learning framework with large language models}. In \bibinfo{booktitle}{\emph{Proceedings of the 32nd ACM International Conference on Multimedia}}. \bibinfo{pages}{8586--8594}.
\newblock


\bibitem[Mu et~al\mbox{.}(2024)]%
        {mu2024clarifygpt}
\bibfield{author}{\bibinfo{person}{Fangwen Mu}, \bibinfo{person}{Lin Shi}, \bibinfo{person}{Song Wang}, \bibinfo{person}{Zhuohao Yu}, \bibinfo{person}{Binquan Zhang}, \bibinfo{person}{ChenXue Wang}, \bibinfo{person}{Shichao Liu}, {and} \bibinfo{person}{Qing Wang}.} \bibinfo{year}{2024}\natexlab{}.
\newblock \showarticletitle{Clarifygpt: A framework for enhancing llm-based code generation via requirements clarification}.
\newblock \bibinfo{journal}{\emph{Proceedings of the ACM on Software Engineering}} \bibinfo{volume}{1}, \bibinfo{number}{FSE} (\bibinfo{year}{2024}), \bibinfo{pages}{2332--2354}.
\newblock


\bibitem[Nasiri and Garraghan(2025)]%
        {nasiri2025edora}
\bibfield{author}{\bibinfo{person}{Hamid Nasiri} {and} \bibinfo{person}{Peter Garraghan}.} \bibinfo{year}{2025}\natexlab{}.
\newblock \showarticletitle{EDoRA: Efficient Weight-Decomposed Low-Rank Adaptation via Singular Value Decomposition}.
\newblock \bibinfo{journal}{\emph{arXiv preprint arXiv:2501.12067}} (\bibinfo{year}{2025}).
\newblock


\bibitem[Nazari et~al\mbox{.}(2024)]%
        {nazari2024entropy}
\bibfield{author}{\bibinfo{person}{Philipp Nazari}, \bibinfo{person}{Oliver Lemke}, \bibinfo{person}{Davide Guidobene}, {and} \bibinfo{person}{Artiom Gesp}.} \bibinfo{year}{2024}\natexlab{}.
\newblock \showarticletitle{Entropy Aware Message Passing in Graph Neural Networks}.
\newblock \bibinfo{journal}{\emph{arXiv preprint arXiv:2403.04636}} (\bibinfo{year}{2024}).
\newblock


\bibitem[Newman et~al\mbox{.}(2020)]%
        {newman2020eos}
\bibfield{author}{\bibinfo{person}{Benjamin Newman}, \bibinfo{person}{John Hewitt}, \bibinfo{person}{Percy Liang}, {and} \bibinfo{person}{Christopher~D Manning}.} \bibinfo{year}{2020}\natexlab{}.
\newblock \showarticletitle{The eos decision and length extrapolation}.
\newblock \bibinfo{journal}{\emph{arXiv preprint arXiv:2010.07174}} (\bibinfo{year}{2020}).
\newblock


\bibitem[Nijkamp et~al\mbox{.}(2022)]%
        {nijkamp2022codegen}
\bibfield{author}{\bibinfo{person}{Erik Nijkamp}, \bibinfo{person}{Bo Pang}, \bibinfo{person}{Hiroaki Hayashi}, \bibinfo{person}{Lifu Tu}, \bibinfo{person}{Huan Wang}, \bibinfo{person}{Yingbo Zhou}, \bibinfo{person}{Silvio Savarese}, {and} \bibinfo{person}{Caiming Xiong}.} \bibinfo{year}{2022}\natexlab{}.
\newblock \showarticletitle{Codegen: An open large language model for code with multi-turn program synthesis}.
\newblock \bibinfo{journal}{\emph{arXiv preprint arXiv:2203.13474}} (\bibinfo{year}{2022}).
\newblock


\bibitem[Pan et~al\mbox{.}(2025)]%
        {Pan2025}
\bibfield{author}{\bibinfo{person}{Ruwei Pan}, \bibinfo{person}{Hongyu Zhang}, {and} \bibinfo{person}{Chao Liu}.} \bibinfo{year}{2025}\natexlab{}.
\newblock \showarticletitle{CodeCoR: An LLM-Based Self-Reflective Multi-Agent Framework for Code Generation}.
\newblock \bibinfo{journal}{\emph{arXiv preprint arXiv:2501.07811}} (\bibinfo{date}{January} \bibinfo{year}{2025}).
\newblock


\bibitem[Pramanick et~al\mbox{.}(2024)]%
        {pramanick2024spiqa}
\bibfield{author}{\bibinfo{person}{Shraman Pramanick}, \bibinfo{person}{Rama Chellappa}, {and} \bibinfo{person}{Subhashini Venugopalan}.} \bibinfo{year}{2024}\natexlab{}.
\newblock \showarticletitle{Spiqa: A dataset for multimodal question answering on scientific papers}.
\newblock \bibinfo{journal}{\emph{arXiv preprint arXiv:2407.09413}} (\bibinfo{year}{2024}).
\newblock


\bibitem[Puangjaktha et~al\mbox{.}(2024)]%
        {puangjaktha2024paper}
\bibfield{author}{\bibinfo{person}{Prahyat Puangjaktha}, \bibinfo{person}{Morakot Choetkiertikul}, {and} \bibinfo{person}{Suppawong Tuarob}.} \bibinfo{year}{2024}\natexlab{}.
\newblock \showarticletitle{“Paper, Meet Code”: A Deep Learning Approach to Linking Scholarly Articles with GitHub Repositories}.
\newblock \bibinfo{journal}{\emph{IEEE Access}} (\bibinfo{year}{2024}).
\newblock


\bibitem[Qian et~al\mbox{.}(2023)]%
        {qian2023communicative}
\bibfield{author}{\bibinfo{person}{Chen Qian}, \bibinfo{person}{Xin Cong}, \bibinfo{person}{Cheng Yang}, \bibinfo{person}{Weize Chen}, \bibinfo{person}{Yusheng Su}, \bibinfo{person}{Juyuan Xu}, \bibinfo{person}{Zhiyuan Liu}, {and} \bibinfo{person}{Maosong Sun}.} \bibinfo{year}{2023}\natexlab{}.
\newblock \showarticletitle{Communicative agents for software development}.
\newblock \bibinfo{journal}{\emph{arXiv preprint arXiv:2307.07924}} \bibinfo{volume}{6}, \bibinfo{number}{3} (\bibinfo{year}{2023}).
\newblock


\bibitem[Raiaan et~al\mbox{.}(2024)]%
        {raiaan2024systematic}
\bibfield{author}{\bibinfo{person}{Mohaimenul Azam~Khan Raiaan}, \bibinfo{person}{Sadman Sakib}, \bibinfo{person}{Nur~Mohammad Fahad}, \bibinfo{person}{Abdullah Al~Mamun}, \bibinfo{person}{Md~Anisur Rahman}, \bibinfo{person}{Swakkhar Shatabda}, {and} \bibinfo{person}{Md~Saddam~Hossain Mukta}.} \bibinfo{year}{2024}\natexlab{}.
\newblock \showarticletitle{A systematic review of hyperparameter optimization techniques in Convolutional Neural Networks}.
\newblock \bibinfo{journal}{\emph{Decision analytics journal}} (\bibinfo{year}{2024}), \bibinfo{pages}{100470}.
\newblock


\bibitem[Seo et~al\mbox{.}(2025)]%
        {seo2025paper2code}
\bibfield{author}{\bibinfo{person}{Minju Seo}, \bibinfo{person}{Jinheon Baek}, \bibinfo{person}{Seongyun Lee}, {and} \bibinfo{person}{Sung~Ju Hwang}.} \bibinfo{year}{2025}\natexlab{}.
\newblock \showarticletitle{Paper2Code: Automating Code Generation from Scientific Papers in Machine Learning}.
\newblock \bibinfo{journal}{\emph{arXiv preprint arXiv:2504.17192}} (\bibinfo{year}{2025}).
\newblock


\bibitem[Song et~al\mbox{.}(2025)]%
        {song2025multidisciplinary}
\bibfield{author}{\bibinfo{person}{Haitao Song}, \bibinfo{person}{Hongyi Xu}, \bibinfo{person}{Zikai Wang}, \bibinfo{person}{Yifan Wang}, {and} \bibinfo{person}{Jiajia Li}.} \bibinfo{year}{2025}\natexlab{}.
\newblock \showarticletitle{A Multidisciplinary Multimodal Aligned Dataset for Academic Data Processing}.
\newblock \bibinfo{journal}{\emph{Scientific Data}} \bibinfo{volume}{12}, \bibinfo{number}{1} (\bibinfo{year}{2025}), \bibinfo{pages}{172}.
\newblock


\bibitem[Starace et~al\mbox{.}(2025)]%
        {starace2025paperbench}
\bibfield{author}{\bibinfo{person}{Giulio Starace}, \bibinfo{person}{Oliver Jaffe}, \bibinfo{person}{Dane Sherburn}, \bibinfo{person}{James Aung}, \bibinfo{person}{Jun~Shern Chan}, \bibinfo{person}{Leon Maksin}, \bibinfo{person}{Rachel Dias}, \bibinfo{person}{Evan Mays}, \bibinfo{person}{Benjamin Kinsella}, \bibinfo{person}{Wyatt Thompson}, {et~al\mbox{.}}} \bibinfo{year}{2025}\natexlab{}.
\newblock \showarticletitle{PaperBench: Evaluating AI's Ability to Replicate AI Research}.
\newblock \bibinfo{journal}{\emph{arXiv preprint arXiv:2504.01848}} (\bibinfo{year}{2025}).
\newblock


\bibitem[Sundar et~al\mbox{.}(2024)]%
        {sundar2024cpapers}
\bibfield{author}{\bibinfo{person}{Anirudh Sundar}, \bibinfo{person}{Jin Xu}, \bibinfo{person}{William Gay}, \bibinfo{person}{Christopher Richardson}, {and} \bibinfo{person}{Larry Heck}.} \bibinfo{year}{2024}\natexlab{}.
\newblock \showarticletitle{cPAPERS: A Dataset of Situated and Multimodal Interactive Conversations in Scientific Papers}.
\newblock \bibinfo{journal}{\emph{Advances in Neural Information Processing Systems}}  \bibinfo{volume}{37} (\bibinfo{year}{2024}), \bibinfo{pages}{66283--66304}.
\newblock


\bibitem[Tao et~al\mbox{.}(2024)]%
        {tao2024magis}
\bibfield{author}{\bibinfo{person}{Wei Tao}, \bibinfo{person}{Yucheng Zhou}, \bibinfo{person}{Yanlin Wang}, \bibinfo{person}{Wenqiang Zhang}, \bibinfo{person}{Hongyu Zhang}, {and} \bibinfo{person}{Yu Cheng}.} \bibinfo{year}{2024}\natexlab{}.
\newblock \showarticletitle{Magis: Llm-based multi-agent framework for github issue resolution}.
\newblock \bibinfo{journal}{\emph{Advances in Neural Information Processing Systems}}  \bibinfo{volume}{37} (\bibinfo{year}{2024}), \bibinfo{pages}{51963--51993}.
\newblock


\bibitem[Thakur et~al\mbox{.}(2024)]%
        {thakur2024verigen}
\bibfield{author}{\bibinfo{person}{Shailja Thakur}, \bibinfo{person}{Baleegh Ahmad}, \bibinfo{person}{Hammond Pearce}, \bibinfo{person}{Benjamin Tan}, \bibinfo{person}{Brendan Dolan-Gavitt}, \bibinfo{person}{Ramesh Karri}, {and} \bibinfo{person}{Siddharth Garg}.} \bibinfo{year}{2024}\natexlab{}.
\newblock \showarticletitle{Verigen: A large language model for verilog code generation}.
\newblock \bibinfo{journal}{\emph{ACM Transactions on Design Automation of Electronic Systems}} \bibinfo{volume}{29}, \bibinfo{number}{3} (\bibinfo{year}{2024}), \bibinfo{pages}{1--31}.
\newblock


\bibitem[Trofimova et~al\mbox{.}(2024)]%
        {trofimova2024coderefine}
\bibfield{author}{\bibinfo{person}{Ekaterina Trofimova}, \bibinfo{person}{Emil Sataev}, {and} \bibinfo{person}{Abhijit~Singh Jowhari}.} \bibinfo{year}{2024}\natexlab{}.
\newblock \showarticletitle{CodeRefine: A Pipeline for Enhancing LLM-Generated Code Implementations of Research Papers}.
\newblock \bibinfo{journal}{\emph{arXiv preprint arXiv:2408.13366}} (\bibinfo{year}{2024}).
\newblock


\bibitem[Wang et~al\mbox{.}(2024)]%
        {wang2024mineru}
\bibfield{author}{\bibinfo{person}{Bin Wang}, \bibinfo{person}{Chao Xu}, \bibinfo{person}{Xiaomeng Zhao}, \bibinfo{person}{Linke Ouyang}, \bibinfo{person}{Fan Wu}, \bibinfo{person}{Zhiyuan Zhao}, \bibinfo{person}{Rui Xu}, \bibinfo{person}{Kaiwen Liu}, \bibinfo{person}{Yuan Qu}, \bibinfo{person}{Fukai Shang}, {et~al\mbox{.}}} \bibinfo{year}{2024}\natexlab{}.
\newblock \showarticletitle{Mineru: An open-source solution for precise document content extraction}.
\newblock \bibinfo{journal}{\emph{arXiv preprint arXiv:2409.18839}} (\bibinfo{year}{2024}).
\newblock


\bibitem[Wang et~al\mbox{.}(2023)]%
        {wang2023codet5+}
\bibfield{author}{\bibinfo{person}{Yue Wang}, \bibinfo{person}{Hung Le}, \bibinfo{person}{Akhilesh~Deepak Gotmare}, \bibinfo{person}{Nghi~DQ Bui}, \bibinfo{person}{Junnan Li}, {and} \bibinfo{person}{Steven~CH Hoi}.} \bibinfo{year}{2023}\natexlab{}.
\newblock \showarticletitle{Codet5+: Open code large language models for code understanding and generation}.
\newblock \bibinfo{journal}{\emph{arXiv preprint arXiv:2305.07922}} (\bibinfo{year}{2023}).
\newblock


\bibitem[Wei et~al\mbox{.}(2024)]%
        {wei2024general}
\bibfield{author}{\bibinfo{person}{Haoran Wei}, \bibinfo{person}{Chenglong Liu}, \bibinfo{person}{Jinyue Chen}, \bibinfo{person}{Jia Wang}, \bibinfo{person}{Lingyu Kong}, \bibinfo{person}{Yanming Xu}, \bibinfo{person}{Zheng Ge}, \bibinfo{person}{Liang Zhao}, \bibinfo{person}{Jianjian Sun}, \bibinfo{person}{Yuang Peng}, {et~al\mbox{.}}} \bibinfo{year}{2024}\natexlab{}.
\newblock \showarticletitle{General ocr theory: Towards ocr-2.0 via a unified end-to-end model}.
\newblock  (\bibinfo{year}{2024}).
\newblock


\bibitem[Wu et~al\mbox{.}(2024)]%
        {wu2024visionllm}
\bibfield{author}{\bibinfo{person}{Jiannan Wu}, \bibinfo{person}{Muyan Zhong}, \bibinfo{person}{Sen Xing}, \bibinfo{person}{Zeqiang Lai}, \bibinfo{person}{Zhaoyang Liu}, \bibinfo{person}{Zhe Chen}, \bibinfo{person}{Wenhai Wang}, \bibinfo{person}{Xizhou Zhu}, \bibinfo{person}{Lewei Lu}, \bibinfo{person}{Tong Lu}, {et~al\mbox{.}}} \bibinfo{year}{2024}\natexlab{}.
\newblock \showarticletitle{Visionllm v2: An end-to-end generalist multimodal large language model for hundreds of vision-language tasks}.
\newblock \bibinfo{journal}{\emph{Advances in Neural Information Processing Systems}}  \bibinfo{volume}{37} (\bibinfo{year}{2024}), \bibinfo{pages}{69925--69975}.
\newblock


\bibitem[Wu et~al\mbox{.}(2023)]%
        {wu2023autogen}
\bibfield{author}{\bibinfo{person}{Qingyun Wu}, \bibinfo{person}{Gagan Bansal}, \bibinfo{person}{Jieyu Zhang}, \bibinfo{person}{Yiran Wu}, \bibinfo{person}{Beibin Li}, \bibinfo{person}{Erkang Zhu}, \bibinfo{person}{Li Jiang}, \bibinfo{person}{Xiaoyun Zhang}, \bibinfo{person}{Shaokun Zhang}, \bibinfo{person}{Jiale Liu}, {et~al\mbox{.}}} \bibinfo{year}{2023}\natexlab{}.
\newblock \showarticletitle{Autogen: Enabling next-gen llm applications via multi-agent conversation}.
\newblock \bibinfo{journal}{\emph{arXiv preprint arXiv:2308.08155}} (\bibinfo{year}{2023}).
\newblock


\bibitem[Xu et~al\mbox{.}(2024a)]%
        {xu2024large}
\bibfield{author}{\bibinfo{person}{Jinglue Xu}, \bibinfo{person}{Jialong Li}, \bibinfo{person}{Zhen Liu}, \bibinfo{person}{Nagar Anthel~Venkatesh Suryanarayanan}, \bibinfo{person}{Guoyuan Zhou}, \bibinfo{person}{Jia Guo}, \bibinfo{person}{Hitoshi Iba}, {and} \bibinfo{person}{Kenji Tei}.} \bibinfo{year}{2024}\natexlab{a}.
\newblock \showarticletitle{Large language models synergize with automated machine learning}.
\newblock \bibinfo{journal}{\emph{arXiv preprint arXiv:2405.03727}} (\bibinfo{year}{2024}).
\newblock


\bibitem[Xu et~al\mbox{.}(2024b)]%
        {xu2024no}
\bibfield{author}{\bibinfo{person}{Minghao Xu}, \bibinfo{person}{Lichuan Xiang}, \bibinfo{person}{Xu Cai}, {and} \bibinfo{person}{Hongkai Wen}.} \bibinfo{year}{2024}\natexlab{b}.
\newblock \showarticletitle{No More Adam: Learning Rate Scaling at Initialization is All You Need}.
\newblock \bibinfo{journal}{\emph{arXiv preprint arXiv:2412.11768}} (\bibinfo{year}{2024}).
\newblock


\bibitem[Yang et~al\mbox{.}(2024)]%
        {yang2024swe}
\bibfield{author}{\bibinfo{person}{John Yang}, \bibinfo{person}{Carlos Jimenez}, \bibinfo{person}{Alexander Wettig}, \bibinfo{person}{Kilian Lieret}, \bibinfo{person}{Shunyu Yao}, \bibinfo{person}{Karthik Narasimhan}, {and} \bibinfo{person}{Ofir Press}.} \bibinfo{year}{2024}\natexlab{}.
\newblock \showarticletitle{Swe-agent: Agent-computer interfaces enable automated software engineering}.
\newblock \bibinfo{journal}{\emph{Advances in Neural Information Processing Systems}}  \bibinfo{volume}{37} (\bibinfo{year}{2024}), \bibinfo{pages}{50528--50652}.
\newblock


\bibitem[Yang et~al\mbox{.}(2023)]%
        {yang2023deep}
\bibfield{author}{\bibinfo{person}{Zezhou Yang}, \bibinfo{person}{Sirong Chen}, \bibinfo{person}{Cuiyun Gao}, \bibinfo{person}{Zhenhao Li}, \bibinfo{person}{Ge Li}, {and} \bibinfo{person}{Michael Lyu}.} \bibinfo{year}{2023}\natexlab{}.
\newblock \showarticletitle{Deep Learning Based Code Generation Methods: Literature Review}.
\newblock \bibinfo{journal}{\emph{arXiv preprint arXiv:2303.01056}} (\bibinfo{year}{2023}).
\newblock


\bibitem[Zhang et~al\mbox{.}(2024)]%
        {zhang2024codeagent}
\bibfield{author}{\bibinfo{person}{Kechi Zhang}, \bibinfo{person}{Jia Li}, \bibinfo{person}{Ge Li}, \bibinfo{person}{Xianjie Shi}, {and} \bibinfo{person}{Zhi Jin}.} \bibinfo{year}{2024}\natexlab{}.
\newblock \showarticletitle{Codeagent: Enhancing code generation with tool-integrated agent systems for real-world repo-level coding challenges}.
\newblock \bibinfo{journal}{\emph{arXiv preprint arXiv:2401.07339}} (\bibinfo{year}{2024}).
\newblock


\bibitem[Zhang et~al\mbox{.}(2023)]%
        {zhang2023planning}
\bibfield{author}{\bibinfo{person}{Shun Zhang}, \bibinfo{person}{Zhenfang Chen}, \bibinfo{person}{Yikang Shen}, \bibinfo{person}{Mingyu Ding}, \bibinfo{person}{Joshua~B Tenenbaum}, {and} \bibinfo{person}{Chuang Gan}.} \bibinfo{year}{2023}\natexlab{}.
\newblock \showarticletitle{Planning with large language models for code generation}.
\newblock \bibinfo{journal}{\emph{arXiv preprint arXiv:2303.05510}} (\bibinfo{year}{2023}).
\newblock


\bibitem[Zheng et~al\mbox{.}(2023)]%
        {zheng2023judging}
\bibfield{author}{\bibinfo{person}{Lianmin Zheng}, \bibinfo{person}{Wei-Lin Chiang}, \bibinfo{person}{Ying Sheng}, \bibinfo{person}{Siyuan Zhuang}, \bibinfo{person}{Zhanghao Wu}, \bibinfo{person}{Yonghao Zhuang}, \bibinfo{person}{Zi Lin}, \bibinfo{person}{Zhuohan Li}, \bibinfo{person}{Dacheng Li}, \bibinfo{person}{Eric Xing}, {et~al\mbox{.}}} \bibinfo{year}{2023}\natexlab{}.
\newblock \showarticletitle{Judging llm-as-a-judge with mt-bench and chatbot arena}.
\newblock \bibinfo{journal}{\emph{Advances in Neural Information Processing Systems}}  \bibinfo{volume}{36} (\bibinfo{year}{2023}), \bibinfo{pages}{46595--46623}.
\newblock


\bibitem[Zhong et~al\mbox{.}(2024)]%
        {zhong2024evaluation}
\bibfield{author}{\bibinfo{person}{Tianyang Zhong}, \bibinfo{person}{Zhengliang Liu}, \bibinfo{person}{Yi Pan}, \bibinfo{person}{Yutong Zhang}, \bibinfo{person}{Yifan Zhou}, \bibinfo{person}{Shizhe Liang}, \bibinfo{person}{Zihao Wu}, \bibinfo{person}{Yanjun Lyu}, \bibinfo{person}{Peng Shu}, \bibinfo{person}{Xiaowei Yu}, {et~al\mbox{.}}} \bibinfo{year}{2024}\natexlab{}.
\newblock \showarticletitle{Evaluation of openai o1: Opportunities and challenges of agi}.
\newblock \bibinfo{journal}{\emph{arXiv preprint arXiv:2409.18486}} (\bibinfo{year}{2024}).
\newblock


\end{thebibliography}

\end{document}